\documentclass[pre,twocolumn,nofootinbib,twoside,showpacs,superscriptaddress,tightenlines,floatfix,longbibliography]{revtex4-2}
\usepackage{dcolumn}
\usepackage{lipsum}
\usepackage{graphicx,amssymb,amsmath,epsf,bm}
\usepackage[utf8x]{inputenc}
\usepackage{tikz}
\usepackage[compat=1.1.0]{tikz-feynman}
\usepackage[T1]{fontenc}
\usepackage{babel}
\usepackage[normalem]{ulem}
\usepackage{color}
\usepackage{mathtools}
\usepackage{dsfont}
\usepackage[dvipsnames]{xcolor}
\usepackage{nicefrac}
\usepackage{upgreek}
\usepackage{cancel}
\tikzfeynmanset{compat=1.1.0}
\usepackage{hyperref}
\hypersetup{
    allcolors=blue,
    colorlinks=true,
    bookmarks=true,
    pdfpagemode=FullScreen,
}
\definecolor{nred}{RGB}{224,0,0}
\definecolor{nblue}{RGB}{28,130,185}
\definecolor{ngreen}{RGB}{38,238,21}
\definecolor{norange}{RGB}{230,120,20}

\usepackage{tikz}
\usepackage{tikz-feynman}
\tikzfeynmanset{compat=1.1.0}
\begin{document}


\title{Conformal Invariance of the large-$N$ limit of the $O(N)$ universality class}

\author{Santiago Cabrera}
\affiliation{Instituto de F\'isica, Facultad de Ciencias, Universidad de la
	Rep\'ublica, Igu\'a 4225, 11400, Montevideo, Uruguay}%

\author{Gonzalo De Polsi}
\affiliation{Instituto de F\'isica, Facultad de Ciencias, Universidad de la
	Rep\'ublica, Igu\'a 4225, 11400, Montevideo, Uruguay}%

\author{Adam Ran\c{c}on }
\affiliation{Univ. Lille, CNRS, UMR 8523 -- PhLAM -- Laboratoire de Physique
des Lasers Atomes et Mol\'ecules, F-59000 Lille, France}
\affiliation{Institut Universitaire de France}

\author{Nicol\'as Wschebor}
\affiliation{Instituto de F\'isica, Facultad de Ingenier\'ia, Universidad de la
	Rep\'ublica, J.H.y Reissig 565, 11000 Montevideo, Uruguay}%

\begin{abstract}
	Conformal symmetry is expected to be realized in many equilibrium statistical mechanical systems at criticality. Although this is certainly true in two-dimensional systems, the three-dimensional case is subtler, and only a few proofs exist, only so in very specific cases. In this work, we give two proofs for the large $N$ limit of the $O(N)$ universality class within the non-perturbative renormalization group framework: one functional, and one vertex-by-vertex in Fourier space. While doing so, we unveil how the theory is structured in order for conformal symmetry to be realized. As a consequence, we shed light on what to expect, on rather general grounds, for a theory to be conformally invariant.
\end{abstract}

\maketitle

\section{Introduction}

Symmetries play a central role in our understanding of quantum and statistical field theories. Among them, scale and conformal invariance occupy a particularly prominent position, as they govern the long-distance behavior of systems at criticality and strongly constrain the structure of correlation functions. While scale invariance is a generic consequence of the existence of renormalization group (RG) fixed points, the question of whether it is automatically enhanced to full conformal invariance has been the subject of longstanding investigation and debate.

In two dimensions, the situation is well understood: under rather mild assumptions such as unitarity and a discrete spectrum of scaling dimensions, scale invariance implies conformal invariance \cite{Polchinski1988,Zamolodchikov1986}. The resulting infinite-dimensional conformal symmetry algebra provides an exceptionally powerful framework, enabling the exact classification of universality classes and the computation of critical exponents \cite{DiFrancesco:1997nk}. In dimensions greater than two, however, the status of conformal invariance at RG fixed points is considerably subtler. Although conformal symmetry is widely believed to emerge in most physically relevant critical systems, general proofs are scarce and non-rigorous (see, for example, \cite{delamotte2016scale,DePolsi2019}), and counterexamples are known when certain assumptions are relaxed (see, for example, \cite{riva2005scale}).

In three dimensions in particular, the issue of scale versus conformal invariance remains only partially resolved. Early arguments based on the structure of the stress-energy tensor suggest that conformal invariance should hold for theories that, when extended to the Minkowskian setting, are unitary field theories at criticality, provided no virial current obstructs the improvement of the stress tensor \cite{Polchinski1988}. Nevertheless, establishing this enhancement in a fully non-perturbative and model-independent manner has proven difficult. As a result, rigorous demonstrations of conformal invariance are limited to specific classes of theories or rely on additional assumptions whose validity is not always transparent.

Among the paradigmatic models of critical phenomena, the $O(N)$-symmetric scalar $\phi^4$ theory plays a central role. It provides a unified description of a wide variety of universality classes, including the Ising, XY, and Heisenberg models, and serves as a testing ground for analytical methods ranging from perturbative RG and the $\epsilon$-expansion to conformal bootstrap techniques \cite{Wilson:1973jj,ZinnJustin:2002ru,Pelissetto02,Poland:2018epd}. In three dimensions, the Wilson-Fisher fixed point of the $\phi^4$ theory is widely believed to be conformally invariant, and this assumption underlies much of the remarkable success of the modern numerical bootstrap program. 

The heuristic reasons why many critical phenomena in equilibrium are expected to be conformally invariant are easy to explain. One can prove, based on the structure of the conformal algebra, that any model invariant under translations, rotations, parity, and dilations is automatically conformally invariant if it does not have any local vector operator (invariant under the internal symmetries of the model that is not a total derivative) with a scaling dimension exactly equal to $d-1$ (where $d$ is the dimension of space). Now, let us consider, for example, the $O(N)$ models for which the upper critical dimension is $d=4$. The most relevant local vector operator that is not a total derivative in $d=4$ has dimension 7 \cite{delamotte2016scale,DePolsi2019}, so at every order of the $\epsilon-$expansion the theory must be conformally invariant since in a finite neighborhood of $d=4$ no vector operator can have scaling dimension $d-1$. Furthermore, to reach the value $d-1=2$ for $d=3$ would require a totally pathological behavior of the perturbation theory in order to bring the dimension from 7 to 2, when going from $d=4$ to $d=3$. Despite these general considerations, a general and rigorous first-principles proof of conformal invariance remains highly nontrivial.

A particularly fruitful limit in which the $O(N)$ model becomes analytically tractable is the large-$N$ limit. In this regime, the theory exhibits a nontrivial interacting fixed point already at leading order, while many complicated dynamical features simplify or reorganize in a controlled manner. The large-$N$ expansion has long been used to compute critical exponents, operator dimensions, and correlation functions, and it provides a natural laboratory to investigate deep structural questions such as the emergence of symmetries at criticality. 

Recently, Pagani and Sonoda \cite{Pagani:2025dtc} showed that in the large-$N$ limit, the energy-momentum tensor of $O(N)$ models has correlation functions that satisfy the requirements formulated by Polchinski \cite{Polchinski1988} to ensure the presence of conformal invariance at the critical point.
Despite this, the mechanism by which conformal invariance is realized at large-$N$ has not been fully clarified from a renormalization group perspective. In fact, the behaviour of the energy-momentum tensor is  
indirect evidence of the presence of conformal invariance and, moreover, it does not explicitly show how the Ward identities (WI) associated with conformal invariance, modified in the presence of the infrared regulator, are satisfied. As we shall see, this is not trivial at all. 

In this work, we address this issue within the framework of the non-perturbative renormalization group (NPRG), also known as the functional renormalization group. This approach allows one to follow the RG flow of the full effective action, interpolating smoothly between microscopic and macroscopic scales, without relying on small coupling expansions. As such, it is particularly well-suited to addressing questions of symmetry realization and symmetry enhancement at fixed points. 

The role of conformal invariance within the NPRG framework has given rise to several works in the literature \cite{Rosten:2014oja,Sonoda:2015pva,delamotte2016scale,Rosten:2016nmc,Rosten:2016zap,Rosten:2017urs,Sonoda2017,DePolsi2019,Balog2020,Delamotte2024,Cabrera2025,Pagani:2025dtc}. These works address various aspects that can be grouped into two categories. The first category includes some general aspects, such as the proof of the existence of conformal invariance in the Ising model or for the O(N) models (with $N=2$, $3,$ and $4$) using the NPRG \cite{delamotte2016scale,DePolsi2019}, through analysis of the spectrum of vector operators in the Ising and $O(N)$ models, the construction of functional representations of the conformal group \cite{Rosten:2014oja,Rosten:2016nmc,Rosten:2016nmc,Rosten:2017urs,Sonoda2017}, and the construction of the energy-momentum tensor in the context of the NPRG \cite{Sonoda:2015pva,Pagani:2025dtc}. A second aspect that has been addressed is the use of conformal invariance to optimize approximation schemes based on the NPRG, such as Derivative Expansion \cite{Balog2020,Delamotte2024,Cabrera2025}.

Focusing on the large-$N$ limit of the $d$-dimensional $O(N)$ model (with $2<d<4$), we provide in the present work two proofs that the RG fixed point governing the critical behavior is indeed conformally invariant. One way to prove this is by showing that conformal WI can be written as a \emph{conformal} operator acting on the solution of the fixed point equation. The other proof goes into the structure of modified conformal transformations WI in the presence of the regulator and verifies that special conformal WI for every vertex are satisfied.

It is worth noting that the explicit construction of an energy-momentum tensor, as presented in \cite{Pagani:2025dtc}, allows us to write the identities associated with Noether’s theorem for isometries, dilations, and special conformal transformations. Once these are integrated, this yields the WI corresponding to those symmetries, implying the presence of conformal invariance for the limit $N\to \infty$ of the $O(N)$ models. This latter aspect was not analyzed in \cite{Pagani:2025dtc}, but it follows naturally from Polchinski’s work \cite{Polchinski1988}, once the energy-momentum tensor is known. In other words, the paper \cite{Pagani:2025dtc} can be viewed as another functional proof of the presence of conformal invariance in this context.

Beyond establishing conformal invariance itself, our analysis reveals in detail how the internal structure of the theory reorganizes along the RG flow to allow this symmetry enhancement to occur. In particular, we show how potentially obstructing contributions are dynamically suppressed at the fixed point, leading to a consistent conformal structure. This perspective not only clarifies the status of conformal symmetry in the large-$N$ limit of $O(N)$ models, but also offers insight into the general conditions under which scale-invariant fixed points are expected to be conformally invariant.

The paper is organized as follows. In Sec.~\ref{sec:NPRG} we briefly review the nonperturbative renormalization group formalism and its implementation for the $O(N)$ scalar theory. In Sec.~\ref{sec:largeN}, we discuss the large $N$ limit and derive the exact flow equations governing the fixed point. Then, in Sec.~\ref{sec:functionalProof} we give the first proof for conformal invariance in a functional approach. We continue in Sec.~\ref{sec:conformal} with the second proof of conformal invariance and a detailed analysis of the underlying inner mechanism. Finally, in Sec.~\ref{sec:discussion} we summarize our findings and discuss their implications for more general theories and for the broader question of conformal invariance at RG fixed points.

\section{Non perturbative renormalization group and the symmetries at criticality}\label{sec:NPRG}

We consider the $O(N)$ model in dimension $2<d<4$, with typical microscopic action
\begin{equation}\label{eq:micAct}
S[\vec{\varphi}]=\int_x\bigg(\frac{1}{2}(\nabla\vec{\varphi})^2+\frac r2\vec{\varphi}^2+\frac{u}{4!}(\vec{\varphi}^2)^2\bigg),
\end{equation}
where $\varphi_a(x)$ is a $N$-component scalar field and $\int_x$ stands for $\int d^dx$. We assume that the theory is regularized in the ultraviolet by a momentum cut-off $\Lambda$. This model displays a continuous phase transition between a disordered phase at large $r$ and an ordered phase at negative enough $r$.

The modern version of Wilson's RG, known as the functional or non-perturbative renormalization group (NPRG), introduces a scale-dependent regulating term into the action $\Delta S_k[\vec{\varphi}]$, which accounts for the procedure of integrating out the rapid modes (typically with wavenumber $q\gtrsim k$), generating a flow between effective actions. To this end, the regulating term must effectively freeze short-wavenumber fluctuations ($q\lesssim k$) in order to leave unaltered this regime. The regulating term is written as a momentum-dependent term quadratic in the fields \cite{Polchinski:1983gv} or, in direct space,
\begin{equation}
	\Delta S_k[\vec{\varphi}]=\frac{1}{2}\int_{x,y} \varphi_a(x) R_{k,ab}(x,y)\varphi_b(y).
\end{equation}
Here and below, Einstein summation convention is employed. The regulating function $R_{k,ab}(x,y)$ is generally taken diagonal in the $O(N)$ index to preserve the $O(N)$ symmetry. Furthermore, it is assumed to be translation and rotation invariant, which leads to $R_{k,ab}(x,y)=\delta_{ab}R_k(x,y)$ with
\begin{equation}\label{eq:regProfile}
\begin{split}
	R_k(x,y)&= Z_k k^{2+d} r(k^2|x-y|^2).
\end{split}
\end{equation}
We will also work with its Fourier transform $R_k(q^2)= Z_k k^2 r(q^2/k^2)$, where $r$ is a function that vanishes faster than any power-law for momentum $q\gtrsim k$, and it behaves as a large mass term for $q\to 0$. Furthermore, it vanishes at $k=0$, such that we recover the model of interest in this limit, and diverges for $k=\Lambda$, corresponding to the initial condition of the flow, where the mean-field approximation becomes exact.

Supplemented with the regulating term, the functional integral takes the form
\begin{equation}\label{eq:functInt}
	e^{W_k[\vec{J}]}=\int \mathcal{D}\vec{\varphi}\,e^{-S[\vec{\varphi}]-\Delta S_k[\vec{\varphi}]+\int_x \vec{J}\cdot\vec{\varphi}}.
\end{equation}
The NPRG is implemented by introducing a generalized Legendre transform of $W_k[\vec{J}]$, namely the \textit{scale-dependent effective action} $\Gamma_k[\vec{\phi}]$, which turns out to be better suited than $W_k$ for approximations going beyond perturbation theory and is defined by the relation:
\begin{equation}\label{eq:EEAdef}
	\Gamma_k[\vec{\phi}]=-W_k[\vec{J}]+\int_x \vec{J}\cdot\vec{\phi}-\frac{1}{2}\int_{x,y} \phi_a(x) R_{k,ab}(x,y)\phi_b(y),
\end{equation}
where $\phi_a(x)=\frac{\delta W_k[\vec{J}]}{\delta J_a(x)}$ is a scale dependent order parameter (somewhat equivalent to a magnetization in magnetic systems).\footnote{Notice that in many expression we suppress the explicit appearance of the scale $k$ in order to ease notation. It should be clear from context, however, what is scale-dependent and what it is not.}

Following standard procedures \cite{Wetterich:1992yh,Ellwanger:1993kk,Morris:1993qb} (for reviews on the topic, see \cite{Berges:2000ew,Delamotte:2007pf,Dupuis:2020fhh}), one can write down an integro-differential functional evolution equation for $\Gamma_k[\vec{\phi}]$
\begin{equation}\label{eq:NPRGEq}
	\partial_t\Gamma_k[\vec{\phi}]=\frac{1}{2}\int_{x,y}\partial_t R_{k,ab}(x,y)G_{k,ab}[x,y],
\end{equation}
where $t=\log(k)$ is the renormalization group time and 
\begin{equation}\label{eq:Prop}
\begin{split}
    	G_{k,ab}[x,y]&=2\frac{\delta \Gamma_k[\vec{\phi}]}{\delta R_{k,ab}(x,y)},\\ 
        &=\bigg(\frac{\delta^2 \Gamma_k[\vec{\phi}]}{\delta \phi_a(x)\delta \phi_b(y)}+R_{k,ab}(x,y)\bigg)^{-1},
\end{split}
\end{equation}
understood in the operator sense, is the full propagator in the presence of the regulator. In the context of $O(N)$ symmetric theories, it proves useful to introduce the $O(N)$ invariant variable
\begin{equation}
\label{eq:defrho}
\rho(x)\equiv\frac{1}{2}\vec\phi^2(x).
\end{equation}
When supplemented with a rescaling of variables in terms of the scale $k$, 
\begin{equation}\label{eq:rescaling}
	\begin{split}
		 x &= k^{-1} \tilde{x}\;,\;\;\; \phi_a (x) =k^{(d-2)/2}Z_k^{-1/2}\tilde{\phi_a}(\tilde{x}),\\
		 \vec{q} &= k \tilde{\vec{q}}\;,\;\;\;\;\;\;\;\;\;\; \rho (x) =k^{d-2}Z_k^{-1}\tilde{\rho}(\tilde{x}),
	\end{split}
\end{equation}
the flow equation becomes:
\begin{equation}\label{eq:WettEqDless}
	\begin{split}
		\partial_t \Gamma_k[\vec{\tilde{\phi}}]=&\int_{\tilde{x}}\frac{\delta \Gamma_k}{\delta \tilde{\phi}_a(\tilde{x})}\big(\tilde{x}^\nu\tilde{\partial}_\nu+\frac{d-2+\eta_k}{2}\big)\tilde\phi_a(\tilde{x})\\+\frac{1}{2}\int_{\tilde{x},\tilde{y}}&\bigg[(d+2-\eta_k)r(|\tilde{x}-\tilde{y}|)\\&+|\tilde{x}-\tilde{y}|r'(|\tilde{x}-\tilde{y}|)\bigg]\tilde{G}_k(\tilde{x},\tilde{y}),
\end{split}\end{equation}
where the running anomalous dimension $\eta_k$ is defined as $\partial_t Z_k=-\eta_k Z_k$. At its critical regime, the physical theory flows toward a fixed point for which $\eta_k^*=\eta$ becomes the anomalous dimension of the field and is related to the scaling dimension of the field in the familiar way $D_\varphi=(d-2+\eta)/2$. 

Since the RG procedure relates to a dilation and re-scaling of the system, it is not surprising that the modified dilation WI is equivalent to the fixed point equation \cite{delamotte2016scale,DePolsi2019}:
\begin{equation}
\label{FPeq}
		\partial_t \Gamma_k[\vec{\tilde{\phi}}]=0.
\end{equation}
Indeed, let us assume the action and the integration measure in the functional integral to be invariant under all conformal transformations, namely, translations, rotations, dilatations, and special conformal transformations (SCT). The corresponding infinitesimal transformations are realized by the following variations of the fields:
\begin{equation}\label{eq:TransfField}
	\begin{split}
		\updelta_{\rm tra}\,\varphi_a(x)&=\epsilon_\mu \partial^x_\mu\varphi_a(x),\\
		\updelta_{\rm rot}\,\varphi_a(x)&=\epsilon_{\mu\nu} \big(x_\mu \partial^x_\nu-x_\nu \partial^x_\mu\big) \varphi_a(x),\\
		\updelta_{\rm dil}\,\varphi_a(x)&=\epsilon \big(D^x+D_\varphi\big) \varphi_a(x),\\
		\updelta_{\rm conf}\, \varphi_a(x)&=\epsilon_\mu \big(K^x_\mu-2x_\mu D_\varphi\big) \varphi_a(x),
	\end{split}
\end{equation}
with $\partial_\mu^x=\frac{\partial}{\partial x_\mu}$, $D^x=x_\mu \partial^x_\mu$ and $K^x_\mu=x^2\partial^x_\mu-2x_\mu x_\nu\partial^x_\nu$.
The presence of the regulator modifies the WI induced by these symmetries, and they generically read
\begin{equation}\label{eq:genericWI}
	\begin{split}
		\int_x \frac{\delta\Gamma_k}{\delta\phi_a(x)}\updelta\phi_a(x)=-\int_{x,y}\frac{\delta\Gamma_k}{\delta R_{k,ab}(x,y)}\updelta R_{k,ab}(x,y).
	\end{split}
\end{equation}
For a rotation and translation invariant regulator, $\delta_{\rm tra}R_k=0=\delta_{\rm rot}R_k$, and the corresponding WI are the usual ones. On the other hand,
\begin{equation}\label{eq:trans_Rk}
	\begin{split}
		\delta_{\rm dil}R_k(x,y)&=(D^x+D^y+D_R)R_k(x,y),\\
        \delta_{\rm conf}R_k(x,y)&=(K^x_\mu+K^y_\mu-(x_\mu+y_\mu)D_R)R_k(x,y),
	\end{split}
\end{equation}
with $D_R=2d-2D_\varphi$, and the modified WI for dilatation reads
\begin{equation}\label{Warddilat}
	\begin{split}
		\int_x &\frac{\delta\Gamma_{k}}{\delta\phi_a(x)} \big(D^x+D_\varphi\big) \phi_a(x)\\&=-\frac{1}{2}\int_{x,y}G_{k,aa}[x,y](D^x+D^y+D_R)R_k(x,y),
	\end{split}
\end{equation}
and that of special conformal transformations
\begin{equation}\label{Wardconformal}
	\begin{split}
		\int_x &\frac{\delta\Gamma_{k}}{\delta\phi_a(x)} \big(K^x_\mu-2x_\mu D_\varphi\big) \phi_a(x)\\&=-\frac{1}{2}\int_{x,y}G_{k,aa}[x,y](K^x_\mu+K^y_\mu-(x_\mu+y_\mu)D_R)R_k(x,y).
	\end{split}
\end{equation}
Noting that for a regulator as in Eq.~\eqref{eq:regProfile} we have $\updelta_{\rm dil} R_k(x,y)=\partial_t R_k(x,y)$, one shows that the modified WI for dilatation is equivalent to the fixed-point equation (Eq.~\eqref{eq:WettEqDless} with the left-hand side equated to zero). We also note for later convenience that in this case, one also has $\updelta_{\rm conf}R_k(x,y)=-(x_\mu+y_\mu)\partial_t R_k(x,y)$.

Following the same procedure as given in \cite{Delamotte2024,Cabrera2025}, one can work out the corresponding expressions for the vertex functions
\begin{equation}
	\begin{aligned}
		\Gamma_{a_1\dots a_n}^{(n)}&(p_1,\dots,p_{n-1};\vec \phi)\equiv\\\int_{x_1,\dots,x_{n-1}}&\Gamma_{a_1\dots 	a_n}^{(n)}(x_1,\dots,x_{n-1},0)e^{i\sum_{j=1}^{n-1}x_j\cdot p_j},
	\end{aligned}
\end{equation}
where $\Gamma_{a_1\dots a_n}^{(n)}(x_1,\dots,x_{n};\vec \phi)=\frac{\delta^n \Gamma_k}{\delta\phi_{a_1}(x_1)\dots\delta\phi_{a_n}(x_n)}\Big|_{\vec\phi(x)\equiv\vec\phi}$ are the vertices in a uniform field and translation WI is already implemented due to momentum conservation of vertices (i.e. in $\Gamma_k^{(n)}$ the implicit momentum satisfies $p_n=-p_1-\dots-p_{n-1}$). Below, for notational simplicity, we omit the uniform field $\vec\phi$ in the vertices and the scale $k$ whenever its role is not relevant.

It should be noted that we are considering the {\it regularized} theory. Consequently, the WI that will be obtained (called modified WI) include terms for dilation and special conformal transformations that express that the regulator term is {\it not} scale invariant. At the same time, unlike the unregulated critical theory, the vertices obtained will be smooth both as a function of distances or wave vectors and of the field. Proceeding in this way, the modified WIs take the form:
\begin{widetext}
\textit{Rotations}
\begin{equation}\label{eq:rotGamman0}
	\mathcal{J}_{\mu\nu}\Gamma^{(n)}_{a_1\dots a_n}(p_1,\dots,p_{n-1})\equiv\left(\sum_{i=1}^{n-1}p_i^{\mu}\frac{\partial}{\partial p_i^{\nu}}-p_i^{\nu}\frac{\partial}{\partial p_i^{\mu}}\right)\Gamma^{(n)}_{a_1\dots a_n}(p_1,\dots,p_{n-1})=0,
\end{equation}
\textit{Dilatations}
\begin{equation}\label{eq:dilGamman0}
	\begin{aligned}
		\mathcal{D}\Gamma^{(n)}_{a_1\dots a_n}(p_1,\dots,p_{n-1})\equiv\left[\left(\sum_{i=1}^{n-1}p_i^{\nu}\frac{\partial}{\partial p_i^{\nu}}\right) \right.&
		-d+nD_{\varphi}\left.\vphantom{\sum_{i=1}^{n-1}}+D_{\varphi}\phi_i\frac{\partial}{\partial \phi_i}\right]\Gamma^{(n)}_{a_1\dots a_n}(p_1,\dots,p_{n-1})\\&=\frac{1}{2}\int_{q}\partial_t R_k(q)G_{bc}(q) H^{(n)}_{ca_1\dots a_nd}(q,p_1,\dots,p_{n-1},-q)G_{db}(q),
	\end{aligned}
\end{equation}
\textit{Special conformal transformations:}
\begin{equation}\label{eq:confGamman0}
	\begin{aligned}
		\mathcal{K}_{\mu}\Gamma^{(n)}_{a_1\dots a_n}(p_1,\dots,p_{n-1})\equiv\sum_{i=1}^{n-1}\left[\vphantom{\sum_{i=1}^{n}}\right.p_i^{\mu}\frac{\partial^2}{\partial p_i^{\nu}\partial p_i^{\nu}}&-2p_i^{\nu}\frac{\partial^2}{\partial p_i^{\nu}\partial p_i^{\mu}}
		-2D_{\varphi}\frac{\partial}{\partial p_i^{\mu}}\left.\vphantom{\sum_{i=1}^{n}}\right]\Gamma^{(n)}_{a_1\dots a_n}(p_1,\dots,p_{n-1}) 
		 \\  -2D_{\varphi}\phi_i \left.\frac{\partial}{\partial r^{\mu}}\Gamma^{(n+1)}_{ia_1\dots a_n}(r,p_1,\dots)\right\vert_{r=0}
		=-\frac{1}{2}\int_{q}&\partial_t R_k(q)G_{bc}(q)\left(\frac{\partial}{\partial q^{\mu}}+\frac{\partial}{\partial q'^{\mu}}\right)
		\left.\left\{H_{ca_1\dots a_n d}^{(n)}(q,p_1,\dots,p_{n-1},q')\right\}\right\vert_{q'=-q}G_{db}(q),
	\end{aligned}
\end{equation}
\end{widetext}
where the $H^{(n)}$ functions include all the diagrammatic contributions which involve vertex functions from $\Gamma^{(2)}$ up to $\Gamma^{(n+2)}$ \cite{Dupuis:2020fhh,Delamotte2024,Cabrera2025}.
It is important to highlight the fact that, as mentioned before Eq.~(\ref{FPeq}), the fixed point equation is equivalent to the modified WI associated to the invariance of the theory under dilations given by Eq.~\eqref{eq:dilGamman0}.

Moreover, as already mentioned, because we are considering the regularized theory, the physically relevant solutions to the fixed-point equation of the RG flow are those with regular vertices. 
In most physical situations, these solutions form a discrete set (see, for example, \cite{Morris:1994ie}). As a consequence, at least in principle, the fixed-point equation, by itself, uniquely determines in each domain of attraction of the flow the possible critical theories for a given set of order parameters and symmetries of the problem. Stated otherwise, the modified WI for dilation invariance has much more information than the simpler {\it unregularized} WI for dilation invariance, whose solutions are by no means unique. Therefore, while the usual WI are recovered in the limit $k\to 0$ of Eqs.~\eqref{eq:confGamman0} and \eqref{eq:dilGamman0}, it is much more convenient to work with the regularized ones.

Unfortunately, one does not know how to solve the modified WI without approximations due to the infinite tower of concatenated equations. Accordingly, in practice, to solve these modified WI, one generally needs some sort of truncation procedure (whether a perturbative expansion, a field expansion, or a momentum expansion) to obtain a closed set of equations. Once approximations are implemented, one typically obtains a discrete set of approximate fixed-point solutions. The difficulty, then, is that once approximations are performed, in many cases special conformal invariance cannot be imposed exactly but only approximately \cite{Balog2020,Cabrera2025}.

A particular case where the infinite tower of equations can be solved exactly is the large-$N$ limit of the $O(N)$ scalar theory. For this case, a closed form for the dilation equations can be derived and, moreover, solved. Additionally, as we will show in the following sections, also for this particular case, the WI becomes much simpler, in particular regarding the form of the $H^{(n)}$ contributions once some manipulations are performed.

\section{The large-$N$ limit of $O(N)$ models} \label{sec:largeN}

The exact solution to systems exhibiting critical behavior, by which we mean an explicit expression for the behavior of macroscopic properties, is generically not accessible except in very peculiar situations. Examples of this are generically restricted to two-dimensional systems or above the upper critical dimension. However, in the large-$N$ limit of $O(N)$ models, the explicit solution is known for all correlation functions in any dimension. In this section we first review the analytical solution for the large-$N$ limit of these models. We then introduce an additional Legendre transform, which removes many ornaments that are not essential to the problem, and rewrite the modified WI for conformal transformations.

\subsection{The large-$N$ limit within the NPRG}

One of the simplifications that arises in the large-$N$ limit of $O(N)$ models is that if at some scale $k_0$ the functional $\Gamma_k[\vec{\phi}]$ can be written as 
\begin{equation}\label{eq:largeNfunct}
	\Gamma_{k}[\vec{\phi}]=
    \hat{\Gamma}_{k}[\rho]+\int_x\frac{(\nabla\vec{\phi})^2}{2},
\end{equation}
with $\rho(x)$ given in Eq.~\eqref{eq:defrho}, then this property is preserved throughout the renormalization group flow \cite{Morris1997} for all subsequent scales $k$.\footnote{Note that this stays true if we instead put the theory on a lattice, or add terms with an arbitrary number of derivatives, as long as they are quadratic in the field.} Note that the initial condition of the flow  (at $k=\Lambda$) is $\Gamma_{k=\Lambda}[\vec{\phi}]=S[\vec{\phi}]$, where $S[\vec{\phi}]$ takes its usual form given by Eq.~(\ref{eq:micAct}). Incidentally, this also satisfies \eqref{eq:largeNfunct}, implying that this functional form is preserved along the flow. 

This functional dependence of the effective action simplifies the form of the vertices significantly. We will consider now and onwards vertex functions evaluated at a uniform field configuration coming from the functional $\hat{\Gamma}_{k}[\rho]$  defined by
\begin{equation}
\hat{\Gamma}^{(n)}(x_1,\dots,x_{n};\rho)=\frac{\delta^n \hat{\Gamma}}{\delta\rho(x_1)\dots\delta\rho(x_n)}\Big|_{\rho(x)\equiv\rho}, 
\end{equation}
(and, as before, their associated Fourier transform).
For example, if (\ref{eq:largeNfunct}) is satisfied, the propagator in Eq.~\eqref{eq:NPRGEq} in Fourier space takes the form:
\begin{equation}
	G_{ab}(q;\rho)=\frac{\phi_a\phi_b}{2\rho}G_{||}(q;\rho)+\bigg(\delta_{ab}-\frac{\phi_a\phi_b}{2\rho}\bigg)G_{\perp}(q;\rho),
\end{equation}
where $G_{||}$ and $G_{\perp}$ are the longitudinal and transverse components given by:
\begin{equation}\label{eq:longProp}
	G_{||}(q;\rho)=\frac{1}{q^2+\hat{\Gamma}^{(1)}(\rho)+2\rho\hat{\Gamma}^{(2)}(q;\rho)+R_k(q^2)},
\end{equation}
and
\begin{equation}\label{eq:longtrans}
	G_{\perp}(q;\rho)=\frac{1}{q^2+\hat{\Gamma}^{(1)}(\rho)+R_k(q^2)},
\end{equation}
respectively.

Similarly, in the large-$N$ limit NPRG flow equations are also simplified. In particular, only the transverse component of the propagator will enter in the diagrammatic contributions to flow equations. For example, consider the flow equation of $\hat{\Gamma}^{(1)}(\rho)$ (again, as stated before, in a uniform field configuration). This will take the form in the large-$N$ limit \cite{Morris1997}:
\begin{equation}\label{eq:largeNfloweq}
	\partial_t\hat{\Gamma}^{(1)}(\rho)=-\frac{N}{2}\partial_\rho\hat{\Gamma}^{(1)}(\rho)\int_q \partial_t R_k(q^2)G^2_\perp(q;\rho),
\end{equation}
where $\int_q$ stands for $\int\frac{d^dq}{(2\pi)^d}$.
Eq.~(\ref{eq:largeNfloweq}) is much simpler than a standard flow equation. In particular, given the form of the transverse component of the propagator, the flow of $\hat{\Gamma}^{(1)}(\rho)$ does not depend on higher order vertices $\hat{\Gamma}^{(n)}$ (with $n>1$). More generally the equation for the vertex $\hat{\Gamma}^{(n)}$ will depend, in the large-$N$ limit, only on vertices with $\hat{\Gamma}^{(m)}$ with $m\le n$ \cite{Blaizot2006}. This triangular structure reduces the infinite tower of coupled vertex equations to decoupled finite sets of equations.

On top of this enormous simplification, the resulting flow equations can be solved analytically \cite{Berges:2000ew}. For example, starting from equation (\ref{eq:largeNfloweq}) and performing the following change of variables 
\begin{equation}\label{eq:varChangerhoW}
	\rho\to W(\rho)\equiv\hat{\Gamma}^{(1)}(\rho), \qquad	\rho=f_k(W),
\end{equation}
one arrives at the equation for $f_k(W)$:
\begin{equation}\label{eq:fk}
	\partial_t f_k(W)=\frac{N}{2}\int_q \partial_t R_k(q^2)G^2(q,W),
\end{equation}
where the transverse propagator, from now on denoted as $G$ instead of $G_\perp$, must be seen as a function of $q$ and $W$, but we omit its dependence on $W$ for shortness of notation:
\begin{equation}
 G(q)=\frac{1}{q^2+W+R_k(q^2)}.
\end{equation}

In Eq.~(\ref{eq:fk}), $k$ and $W$ are independent variables, so this equation can be solved explicitly as:
\begin{equation}\label{eq:gapeqF}
\rho=f_k(W)=f_\Lambda(W)-\frac{N}2I_k^{(1)}(W),
\end{equation}
where we introduced the integral 
\begin{equation}\label{eq:I1}
	I_{k}^{(1)}(W)\equiv \int_q\Big\lbrace\frac{1}{q^2+W+R_k(q^2)}-\frac{1}{q^2+W+R_{\Lambda}(q^2)}\Big\rbrace.
\end{equation}
Here we exceptionally made explicit the $k-$dependence.
Note that $I_k^{(1)}$ 
has a strong dependence on the UV scale $\Lambda$, which we take very large. 
It is therefore convenient to absorb  
such a strong UV dependence
into $f_\Lambda$. To do so, let us first determine $f_\Lambda(W)$, for which we must specify the initial condition of the flow. Starting from the Ginzburg-Landau microscopic action (\ref{eq:micAct}), we obtain
\begin{equation}
\label{GLinitialcond}
f_\Lambda(W)=3(W- r)/u.
\end{equation}
The condition to reach a critical point is that when $\rho=0$, one must have $W=0$ if $k\to 0$. That is, at the critical temperature:
\begin{equation}\label{eq:gapeqF}
0=f_\Lambda(W=0)-\frac{N}2{I}^{(1)}_{k=0}(W=0),
\end{equation}
that for the initial condition (\ref{GLinitialcond}) gives
\begin{equation}\label{eq:gapeqF}
	0=-3 r_c/u-\frac{N}2\int_q \Big[\frac{1}{q^2}-\frac{1}{q^2+R_{\Lambda}(q^2)}\Big].
\end{equation}
Then we can subtract the equation for the critical case to obtain the gap equation that now reads:
\begin{equation}
\rho = f_R(W)-\frac{N}2I^{(1)}_{R,k}(W),
\end{equation}
where
\begin{equation}
	{I}_{R,k}^{(1)}(W)\equiv \int_q\Big\{\frac{1}{q^2+W+R_k(q^2)}-\frac1{q^2}\Big\}
\end{equation}
and
\begin{equation}
f_R(W)=3(W- \delta r)/u,
\end{equation}
with $\delta r=r-r_c$ with $r_c$ a UV-sensitive constant. 
One shows that $\delta r=0$ corresponds to the critical point and that the correlation length scales as $\xi\propto (\delta r/u)^{-\nu}$ with $\nu=1/(d-2)$ in the large $N$ limit \cite{Moshe:2003xn}.
Observe that for $d<4$ and for $\Lambda^2\gg W$, the strong dependence on the UV scale is suppressed because
\begin{equation}
\int_q\Big[\frac{1}{q^2+W+R_{\Lambda}(q^2)}-\frac{1}{q^2+R_{\Lambda}(q^2)}\Big]=\mathcal{O}(\Lambda^{d-4}).
\end{equation}
The scaling regime corresponds to field configurations and scales such that the physics is described by the RG fixed point, i.e. such that we are probing scales much larger than the Ginzburg length $\ell_G\propto u^{1/(d-4)}$ and much smaller than the correlation length $\xi$. This corresponds to scales such that $f_\Lambda(W)\ll I^{(1)}_{R,k}(W)$.\footnote{Another way to obtain the scaling limit is to consider that close to the fixed point $\rho\propto k^{d-2}$ and $W\propto k^2$. In units of $k$, $f_\Lambda$ is of order ${\rm max}((k\ell_G)^{4-d},(k\xi)^{2-d})$ while $I^{(1)}_{R,k}(W)$ is of order one. Thus $f_{
\Lambda}$ is negligible in the scaling limit.} Since only $f_\Lambda$ depends on the microscopic details of the model, in this regime the physics becomes universal and the gap equation becomes~\footnote{While it looks different from the fixed point solution given in \cite{knorrExactSolutionsResidual2021}, it is in fact equivalent.}
\begin{equation}\label{eq:rhoFP}
\rho  = -\frac{N}2I^{(1)}_{R,k}(W).
\end{equation}

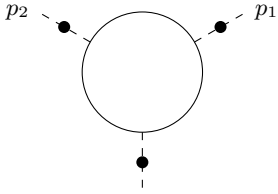
\begin{figure}
	\begin{tikzpicture}
	\begin{feynman}
		\def\R{.8}		
		\pgfmathsetmacro{\RM}{2*\R}
        \pgfmathsetmacro{\rmed}{1.5*\R}
		\vertex (a) at (0,0);
		
		\draw (0,0) circle (\R);
		
		\draw[dashed] (30:\R) -- (30:\RM) node[pos=1, right] {$p_1$};
		\draw[dashed] (-90:\R) -- (-90:\RM)node[pos=1, below] {};
		\draw[dashed] (150:\R) -- (150:\RM)node[pos=1, left] {$p_2$};

        \fill (30:\rmed) circle(0.1*\R);
        \fill (-90:\rmed) circle(0.1*\R);
        \fill (150:\rmed) circle(0.1*\R);
	\end{feynman}
	\end{tikzpicture}
	\caption{\label{Gammahat3} Diagrammatic representation of the $\hat \Gamma^{(3)}$ vertex. Plain lines represent propagators. Integration over the loop is assumed. Dashed lines with a dot represent $\hat \Gamma^{(2)}$ vertices. The loop part is equal to $I^{(3)}$.}
\end{figure}

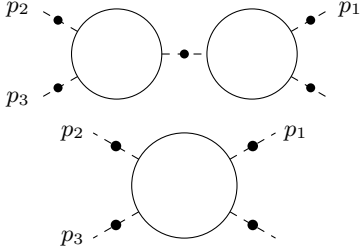
\begin{figure}
\begin{tikzpicture}
\begin{feynman}

\def\R{0.6}

\pgfmathsetmacro{\RM}{2*\R}
\pgfmathsetmacro{\rc}{0.2*\R}
\pgfmathsetmacro{\Rq}{1.2*\R}
\pgfmathsetmacro{\Rp}{1.4*\R}
\pgfmathsetmacro{\rmed}{1.5*\R}

\vertex (a) at (0,0);
\vertex (b) at (3*\R,0);

\draw (0,0) circle (\R);
\draw (b) circle (\R);


\draw[dashed] ($(3*\R,0)+(30:\R)$) -- ($(3*\R,0)+(30:\RM)$) node[right] {$p_1$};
\draw[dashed] ($(3*\R,0)+(-30:\R)$) -- ($(3*\R,0)+(-30:\RM)$);
\draw[dashed] (0:\R) -- (0:2*\R);
\draw[dashed] (150:\R) -- (150:\RM)node[pos=1, left] {$p_2$};
\draw[dashed] (210:\R) -- (210:\RM)node[pos=1,left]{$p_3$};

\fill (150:\rmed) circle(0.1*\R);
\fill (210:\rmed) circle(0.1*\R);
\fill (0:\rmed) circle(0.1*\R);
\fill ($(3*\R,0)+(30:\rmed)$) circle(0.1*\R);
\fill ($(3*\R,0)+(-30:\rmed)$) circle(0.1*\R);

\end{feynman}
\end{tikzpicture}

\begin{tikzpicture}
\begin{feynman}

\def\R{.7}

\pgfmathsetmacro{\RM}{2*\R}
\pgfmathsetmacro{\rmed}{1.5*\R}

\vertex (a) at (0,0);

\draw (0,0) circle (\R);

\draw[dashed] (30:\R) -- (30:\RM) node[pos=1, right] {$p_1$};
\draw[dashed] (-30:\R) -- (-30:\RM)node[pos=1, right] {};
\draw[dashed] (150:\R) -- (150:\RM)node[pos=1, left] {$p_2$};
\draw[dashed] (210:\R) -- (210:\RM)node[pos=1,left]{$p_3$};

\fill (30:\rmed) circle(0.1*\R);
\fill (-30:\rmed) circle(0.1*\R);
\fill (150:\rmed) circle(0.1*\R);
\fill (210:\rmed) circle(0.1*\R);
\end{feynman}
\end{tikzpicture}
\caption{\label{Gammahat4} Diagrammatic representation of two contributions to $\hat \Gamma^{(4)}$ vertex. Plain lines represent propagators. Integration over the loop is assumed. Dashed lines with a dot represent $\hat \Gamma^{(2)}$ vertices. The loop parts are equal to $I^{(3)}$ or $I^{(4)}$. Signs and factors $N$ are omitted (see Eq.~\eqref{eq:Gamma4largeN}). For each diagram, there are two possible extra permutations of momenta.}
\end{figure}

A similar procedure, and using the microscopic action in Eq.~\eqref{eq:micAct}, for the two point function leads to the solution \cite{Blaizot2006}:
\begin{equation}\label{eq:twopointLargeN}
	\hat{\Gamma}^{(2)}(p;W)=\left(\frac{3}{u}+\frac{N}{2}I^{(2)}(p;W)\right)^{-1},
\end{equation}
where we introduced the notation used in the following:
\begin{equation}\label{eq:InDef}
	I^{(n)}(p_1,\dots,p_{n-1};W)\equiv\int_q \prod_{i=0}^{n}G(Q_i;W),
\end{equation}
where we introduced the notation $Q_i=q+\sum_{j=1}^i p_j$ and $Q_0=q$.
Note that $I^{(n\geq2)}$ are finite in the UV for $d<4$ and, consequently, there is no need of subtraction at the scale $\Lambda$ in those expressions.
Notice also that in the definition (\ref{eq:InDef}) there is no explicit reference to $p_n$ in the left-hand side because translation invariance implies that ${p_n=-(p_1+\dots+p_{n-1})}$. One can continue this procedure (see \cite{Blaizot2006}) to obtain all vertex functions.
Here, we give the expressions for the three and four-point functions (omitting the dependence on $W$): 
\begin{equation}\label{eq:Gamma3largeN}
		\hat{\Gamma}^{(3)}(p_1,p_2)=N\hat{\Gamma}^{(2)}(p_1)\hat{\Gamma}^{(2)}(p_2)\hat{\Gamma}^{(2)}(p_3)I^{(3)}(p_1,p_2),
\end{equation}

\begin{equation}\label{eq:Gamma4largeN}
	\begin{split}
		\hat{\Gamma}^{(4)}(p_1,p_2,p_3)&=N\hat{\Gamma}^{(2)}(p_1)\hat{\Gamma}^{(2)}(p_2)\hat{\Gamma}^{(2)}(p_3)\hat{\Gamma}^{(2)}(p_4)\\\bigg\lbrace \Big(NI^{(3)}&(p_1,p_2)\hat{\Gamma}^{(2)}(p_1+p_2)I^{(3)}(p_3,p_4)\\&-I^{(4)}(p_1,p_2,p_3)\Big)+\text{2 perms.}\bigg\rbrace,
	\end{split}
\end{equation}
where $p_n$ in $\hat{\Gamma}^{(n)}$ is used as a shorthand for ${p_n=-(p_1+\dots+p_{n-1})}$ and to highlight the symmetric role of external momenta.
These expressions can be represented diagrammatically as in Figs.~\ref{Gammahat3},\ref{Gammahat4}. For higher vertices, the straightforward generalizations can be represented diagrammatically by writing the most general connected tree-level diagram where lines are associated to $\hat{\Gamma}^{(2)}$ and vertices to sums of permutations of $I^{(n)}$.

For these vertices, the scaling limit corresponds to momentum scales and field such that $3/u$ is negligible compared to $I^{(2)}$ in Eq.~\eqref{eq:twopointLargeN} and, at the same time, a correlation length much bigger than all other length scales in the problem. This corresponds to $\xi^{-1}\ll k,p_i,W^{1/2}\ll \ell_G^{-1}$.

\subsection{A convenient reformulation}

The core of large-$N$ solutions is the $I^{(n)}$ functions. They correspond to one-loop diagrams in contrast with the expressions of $\hat{\Gamma}^{(n)}$ vertices that include arbitrary high orders in loop expansions (see, for example, expressions given by Eqs.~(\ref{eq:twopointLargeN},\ref{eq:Gamma3largeN},\ref{eq:Gamma4largeN})). One would like to reduce the proof of conformal invariance to the structure of $I^{(n)}$ functions. However, at first sight, the expression of $\hat{\Gamma}^{(n)}$ in terms of $I^{(n)}$ is not trivial. This implies that plugging $\hat{\Gamma}^{(n)}$ vertices into the modified WI coming from conformal symmetry would lead to some cumbersome and interrelated expressions which are very involved. However, as stated before, the $\hat{\Gamma}^{(n)}$ vertices can be expressed in terms of tree-level diagrams where the various interaction vertices and propagators are extracted from the $I^{(n)}$ functions. As usual, if a set of correlation functions can be expressed with a tree-level relation to some set of vertices, the generating functionals to those sets of correlation and vertex functions are related by a Legendre transformation. This suggests a workaround by considering the following transformation \cite{Sonoda:2023ohb}:
\begin{equation}\label{eq:gammaDef}
\gamma[\sigma]\equiv\int_x\Big\lbrace\sigma(x)\rho(x)\Big\rbrace-\hat{\Gamma}[\rho].
\end{equation}
This transformation is introduced because directly evaluating the vertices of $\hat{\Gamma}$ is analytically prohibitive; transitioning to $\gamma$ isolates the one-loop integrals ($I^{(n)}$), allowing us to decouple the momentum channels later in the proof.

This immediately implies that the new variable $\sigma$ is nothing but:
\begin{equation}
	\sigma(x)=\hat{\Gamma}^{(1)}[x;\rho],
\end{equation}
which at uniform field means that $\sigma=W$. The converse relationship is
\begin{equation}
	\rho(x)=\gamma^{(1)}[x;\sigma].
\end{equation}

The flow equation of $\gamma[\sigma]$ is obtained from 
\begin{equation}
\begin{split}
 	\partial_t \gamma[\sigma] &= -\partial_t \hat{\Gamma}[\rho],\\
   &= -\frac{N}{2}\int_{x,y}\partial_t R_k(x,y)\Big\lbrace\sigma-\nabla^2+R_k\Big\rbrace^{-1}_{x,y}.
\end{split}
\label{eq:flow_gamma}
\end{equation}
Since now $\sigma$ and $k$ are independent variables, only $R_k$ depends on $k$ on the right-hand side. Accordingly, the flow equation can be integrated trivially
\begin{equation}
\begin{split}
 	 \gamma[\sigma] &= \gamma_{k=\Lambda}[\sigma]
    -\frac{N}{2}\int_{x}\Big\lbrace\log\Big[\sigma-\nabla^2+R_k\Big]\\
    &-\log\Big[\sigma-\nabla^2+R_\Lambda\Big]\Big\rbrace_{x,x}.
\end{split}
\end{equation}
For the microscopic action (\ref{eq:micAct}), up to a constant, $\gamma_{k=\Lambda}[\sigma]=\frac{3}{2u}\int_x(\sigma(x)-r)^2$. The functional version of the gap equation then reads
\begin{equation}
\begin{split}
 	 \rho(x)=\gamma^{(1)}[x;\sigma] &= \frac{3}{u}(\sigma(x)-r)
    -\frac{N}{2}\Big\lbrace\Big[\sigma-\nabla^2+R_k\Big]^{-1}\\
    &-\Big[\sigma-\nabla^2+R_{\Lambda}\Big]^{-1}\Big\rbrace_{x,x}.
\end{split}
\end{equation}
This equation has again an important UV sensitivity. As before,
for $d<4$ this sensitivity in the UV scale $\Lambda$ can be absorbed in a renormalization of $r$. Introducing the free propagator $G_0(x,y)$, the Fourier transform of $1/q^2$, and the full propagator $G[x,y;\sigma]$, such that:
\begin{equation}
    \int_z \left\{\delta(x-z)(\sigma(z)-\nabla_z^2)+R_k(x,z)\right\}G[z,y;\sigma]=\delta(x-y),
\end{equation}
the gap equation can be recast as
\begin{equation}
\begin{split}
 	 \gamma^{(1)}[x;\sigma] &= \frac{3}{u}(\sigma(x)-\delta r)
    -\frac{N}{2}(G[x,x;\sigma]-G_0(x,x)).
\end{split}
\end{equation}
In this functional form, the scaling limit corresponds to field configurations such that $\frac{3}{u}(\sigma(x)-\delta r)$ is negligible compared to $-\frac{N}{2}(G[x,x;\sigma]-G_0(x,x))$, leading to the scaling form
\begin{equation}
    \gamma^{(1)}[x;\sigma]=-\frac{N}{2}\bigl(G[x,x]-G_{0}(x,x)\bigr).
    \label{eq:scal_gamma}
\end{equation}
This is the functional equivalent of Eq.~\eqref{eq:rhoFP}.

From the functional $\gamma[\sigma]$, one can define, similarly as before, the set of vertices
$\gamma^{(n)}(x_1,\dots,x_{n};\sigma)=\frac{\delta^n \gamma}{\delta\sigma(x_1)\dots\delta\sigma(x_n)}\Big|_{\sigma(x)\equiv\sigma}$,\footnote{We will use square brackets for vertices that are functionals (i.e. that depends on the function $\sigma(x)$), e.g. $\gamma^{(n)}[x_1,\dots,x_{n};\sigma]$, and parentheses for vertices that are functions (i.e. that depends on a constant configuration $\sigma$), e.g. $\gamma^{(n)}(x_1,\dots,x_{n};\sigma)$. In particular, $\gamma^{(n)}(x_1,\dots,x_{n};\sigma)=\gamma^{(n)}[x_1,\dots,x_{n};\sigma(x) \equiv \sigma]$. We will often omit the dependence on $\sigma$, and the brackets or parentheses will signify the type of vertex we are considering.} and the associated Fourier transform. We will often omit to write the $\sigma$-dependence of vertices explicitly from now on.
One can then trivially show that
\begin{equation}
	\delta(x-y)=\int_z\hat{\Gamma}^{(2)}(x,z)\gamma^{(2)}(z,y),
\end{equation}
and, more generally, one can obtain $\hat{\Gamma}^{(n)}$ vertices in terms of $\gamma^{(m)}$ vertices. For example, one has expressions such as:
\begin{widetext}
	\begin{equation}
		\hat{\Gamma}^{(3)}(x,y,u)=-\int_{z_1,z_2,z_3}\gamma^{(3)}(z_1,z_2,z_3)\hat{\Gamma}^{(2)}(z_1,x)\hat{\Gamma}^{(2)}(z_2,y)\hat{\Gamma}^{(2)}(z_3,u).
	\end{equation}
More generally, these relations express the $\hat{\Gamma}^{(n)}$ vertices as tree-level diagrams where vertices and propagators are read from $\gamma$. In fact, all the previous expressions can be considered not only for uniform fields but even functionally. However, if the various expressions are evaluated in a uniform field, one can exploit translational invariance that becomes particularly simple in Fourier space. Indeed, when performing a Fourier transform, one immediately finds, in connection with Eqs.~(\ref{eq:twopointLargeN}-\ref{eq:Gamma4largeN}), that in the scaling limit it is
	\begin{equation}
		\label{eq:formgammas}
		\gamma^{(n)}\left(p_1,p_2,\cdots,p_{n-1};\sigma\right)=(-1)^{n}N\sum_{P\{p_i\}}I^{(n)}\left(p_{P(1)},\cdots,p_{P(n-1)};\sigma\right).
	\end{equation}
\end{widetext}
In this expression, the integrals $I^{(n)}$ are the ones appearing in Eq.~\eqref{eq:InDef}, with $W$ replaced by $\sigma$, while $P\{p_i\}$ refers to all the possible permutations of the explicit $n-1$ external momenta yielding independent ``channels''. Here we call a channel a permutation of the $n-1$ external momenta that gives different values for $I^{(n)}\left(p_{P(1)},\cdots,p_{P(n-1)};\sigma\right)$. For example, in the $n=4$ case configurations $(p_1,p_2,p_3)$ and $(p_3,p_2,p_1)$ are equivalent and belong to the same channel. 
To sum up: the Legendre transform (\ref{eq:gammaDef}) implements concretely the idea of expressing all large-$N$ vertices in terms of the one-loop integrals $I^{(n)}$. 

\subsection{Outline of the proof}

Let us now give the proof that critical models belonging to the $O(N)$ universality class in the large $N$ limit are conformally invariant at large scales. The proof is based on two steps. 

The first is the most technical and takes the remainder of the manuscript. It amounts to showing that in the scaling regime, the exact solution in the large $N$ limit obtained above obeys the modified WI of conformal invariance: translations, rotations, dilatations, and SCT. The first two are trivial, since the microscopic theory is translation and rotation-invariant, as is the regulator function. Proving the modified WI for SCT is the most arduous task.

Now, as explained in Sec.~\ref{sec:NPRG}, the dilation WI is equivalent to the fixed point equation of the NPRG. Therefore, the exact solution in the scaling regime is nothing but the fixed point solution to the NPRG flow equation, corresponding here to the Wilson-Fisher (WF) fixed point in the large $N$ limit. Stated otherwise, the WF fixed-point solution is conformally invariant. However, it is important to keep in mind that generic microscopic models in the $O(N)$ universality class are neither scale- nor conformal invariant.

Second, one can discuss how scale and conformal symmetry emerge in models that at the microscopic level do not have those symmetries. Indeed, by definition, all models in the $O(N)$ universality class in the large-$N$ limit will have a flow that brings them to the WF fixed point when fine-tuned to criticality. Since the WF fixed point is conformally invariant, all models belonging to the universality class are.

Thus, the proof is independent of the microscopic action, which might be defined on a lattice, or with an arbitrary potential, as long as it is in the basin of attraction of the fixed point. This includes, in particular, the $O(N)$ non-linear sigma model. Of course, it is a priori very hard to know without an explicit calculation whether a specific model does belong to the universality class. This, however, does not lower the generality of the proof.

Considering the functional $\gamma[\sigma]$ instead of $\Gamma[\vec{\phi}]$ or $\hat{\Gamma}[\rho]$ simplifies the proof greatly, both at the functional level and vertex-by-vertex. In the latter case, one only has to focus on the $I^{(n)}$ structure that appears in the $n$-point function and not on $I^{(m)}$ with $m<n$.
 We show in App.~\ref{app:WI} how the modified WI of $\Gamma$ translates into modified WI for $\gamma$.

The next section gives a functional proof of the first step of the proof, while the following one gives a vertex-by-vertex proof.

\section{Functional proof}\label{sec:functionalProof}

In the present section, we show functionally that the scaling limit of the vertex functional $\gamma^{(1)}[x]$, obtained from the exact solution of the quartic potential, obeys the four modified WI associated to conformal invariance. Concerning the modified WI associated to translations and rotations, we only note that since the bare theory is both translation and rotation invariant, as is the regulator function, these two WI will be automatically satisfied.

\begin{widetext}
As stated above, the modified dilatation WI is equivalent to the fixed point equation. For the vertex $\gamma^{(1)}[x]$, it reads 
\begin{equation}
 \mathcal D  \gamma^{(1)}[x]
\stackrel{?}{=} \frac{N}{2} \int_{y,y'} G[x,y] \, \partial_t R_k(y,y') \, G[y',x],
\label{eq:dil_WI_gamma}
\end{equation}
with $G^{-1}[x,y]= -\nabla^2\delta(x-y)+\delta(x-y)\sigma(x)+R_k(x,y)$ and $\mathcal D$ the dilatation generator (at fixed regulator) for $\gamma^{(1)}$,
\begin{equation}
   \mathcal D=  D^x + 2D_\varphi
- \int_{y} \Big((D^y+d-2D_\varphi)\sigma(y)\Big)\frac{\delta}{\delta \sigma(y)},
\label{eq:dil_gamma}
\end{equation}
where $d-2D_\varphi$ is the scaling dimension of $\sigma(x)$ and we recall that $D^y=y_\nu\partial_\nu^y$. Note that $d-2D_\varphi=2$ in large $N$, and $\sigma$ has thus the scaling dimension of a mass. To show that it is satisfied, we use that, as shown in App.~\ref{app:identities},
\begin{equation}
    \int_{y,y'} G[x,y] \, \partial_t R_k(y,y') \, G[y',x]=-\mathcal D
\bigl(G[x,x]-G_{0}(x,x)\bigr),
\label{eq:tadpole}
\end{equation}
with $\mathcal D$ given in Eq.~\eqref{eq:dil_gamma}. This, in turn, implies that the modified WI for dilatation is obeyed, provided
\begin{equation}
    \mathcal D \gamma^{(1)}[x]\stackrel{?}{=}\mathcal D\left(-\frac{N}{2}\bigl(G[x,x]-G_{0}(x,x)\bigr)\right).
\end{equation}
From the exact solution in the scaling limit, Eq.~\eqref{eq:scal_gamma}, this equation is satisfied, proving dilatation invariance.

To prove that the modified WI for SCT is also satisfied, we need to show that  
\begin{equation}
\mathcal{K}_\mu\,\gamma^{(1)}[x]\stackrel{?}{=} -\frac{N}{2}\int_{y,z}G[x,y]G[z,x]\,\Bigl( K_\mu^y + K_\mu^z - (2d-2D_\varphi)
(y_\mu+z_\mu)\Bigr)R_k(y,z)\,,
\label{eq:WI_conf_gamma}
\end{equation}
where 
\begin{equation}
\begin{split}
\mathcal{K}_\mu&=K_\mu^x - 4D_\varphi x_\mu- \int_z \Big(\left(K_\mu^z-2(d-2D_\varphi)z_\mu\right)\sigma(z)\Big)\,\frac{\delta}{\delta\sigma(z)}
\end{split}
\end{equation}
is the special conformal generator (at fixed regulator) for $\gamma^{(1)}$ and we recall that $K_\mu^x=x^2\partial_\mu^x-2x_\mu x_\nu\partial_\nu^x$.

To prove this, we use the fact, shown also in App.~\ref{app:identities}, that
\begin{equation}\label{eq:id_conf}
    \int_{y,z}\,G[x,y]\,\Bigl( K_\mu^y + K_\mu^z - (2d-2D_\varphi)(y_\mu+z_\mu)\Bigr)R_k(y,z)\,G[z,x]=-\mathcal{K}_\mu\Bigl(G[x,x]-G_0(x,x)\Bigr).
\end{equation}
This allows for recasting Eq.~\eqref{eq:WI_conf_gamma} as
\begin{equation}
    \mathcal{K}_\mu\,\gamma^{(1)}[x]\stackrel{?}{=}-\mathcal{K}_\mu\frac N2\Bigl(G[x,x]-G_0(x,x)\Bigr),
\end{equation}
which is true by Eq.~\eqref{eq:scal_gamma}. We have thus proved functionally that at criticality the $O(N)$ model in the large $N$ limit is conformally invariant.  
\end{widetext}

A few comments are in order. First, we note that the proof of both identities Eqs.~\eqref{eq:tadpole} and \eqref{eq:id_conf} relies on the fact that the anomalous dimension vanishes, which allows for exact cancellations of terms when performing integration by parts. With this approach, we thus cannot discuss other conformally invariant fixed points that would have a non-zero anomalous dimension. Furthermore, one could argue that there might be additional solutions to Eq.~\eqref{eq:dil_WI_gamma} (with $\eta=0$), i.e. additional scale-invariant non-trivial theories. These solutions would, however, be non-analytic in the field or in derivatives. Indeed, assume the existence of a ``homogeneous'' solution $\gamma^{(1)}_h[x]$, such that $\mathcal D \gamma^{(1)}_h[x]=0$. It corresponds to operators of the schematic form $\gamma_h[\sigma]=\int_x \partial^\alpha \sigma(x)^\beta$, implying $\alpha$ derivatives and $\beta$ fields, such that it is scale invariant under $x\to s x$, $\sigma(x)\to s^{d-2D_\phi}\sigma(s x)$.  This implies $\alpha+2\beta=d$. For $\gamma_h$ to be regular and invariant under parity, we need that $\alpha=2m$ and $\beta=n$, with $m$ and $n$ positive integers. We thus obtain the constraint $m+n=d/2$, and since $d/2$ is not an integer, this constraint cannot be satisfied. Therefore, the WF fixed point is the only analytic scale-invariant non-trivial theory in large $N$. Note, however, that non-analytic fixed point solutions are known to exist in the large $N$ limit \cite{Yabunaka17,Yabunaka:2018mju,Yabunaka23}. Whether such solutions could also be conformally invariant is out of the scope of this work.

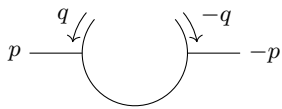
\begin{figure}[t!]
	\begin{tikzpicture}
		
		\def\R{0.7}
		
		\pgfmathsetmacro{\RM}{2*\R}
		\pgfmathsetmacro{\rc}{0.2*\R}
		\pgfmathsetmacro{\Rq}{1.2*\R}
		
		\draw (0,0) ++(40:\R) arc[start angle=40, end angle=-220, radius=\R];
		
		\draw (0:\R) -- (0:\RM) node[pos=1, right] {$-p$};
		\draw (180:\R) -- (180:\RM)node[pos=1, left] {$p$};
		
		\draw[->]
		(0,0) ++(-220:\Rq)
		arc[start angle=-220, end angle=-190, radius=\Rq]
		node[midway, left, yshift=4pt] {$q$};
		
		\draw[->]
		(0,0) ++(40:\Rq)
		arc[start angle=40, end angle=10, radius=\Rq]
		node[midway, right, yshift=4pt] {$-q$};
		
\end{tikzpicture}	\caption{\label{fig:diagramsH2} Only diagram contributing to $H^{(2)}$, where the continuous internal lines correspond to propagators given in Eq.~\eqref{eq:propagator} and vertices are just $1$.}
\end{figure}

\begin{figure}[t!]
	\begin{tikzpicture}
		
		\def\R{.7}
		
		\pgfmathsetmacro{\RM}{2*\R}
		\pgfmathsetmacro{\Rq}{1.2*\R}
		
		\draw (0,0) ++(40:\R) arc[start angle=40, end angle=-220, radius=\R];
		
		\draw (0:\R) -- (0:\RM) node[pos=1, right] {$p_1$};
		\draw (180:\R) -- (180:\RM)node[pos=1, left] {};
		\draw (270:\R) -- (270:\RM)node[pos=1, below] {$p_2$};
		
		\draw[->]
		(0,0) ++(-220:\Rq)
		arc[start angle=-220, end angle=-190, radius=\Rq]
		node[midway, left, yshift=4pt] {$q$};
		
		\draw[->]
		(0,0) ++(40:\Rq)
		arc[start angle=40, end angle=10, radius=\Rq]
		node[midway, right, yshift=4pt] {$-q$};
	
	\end{tikzpicture}
	\begin{tikzpicture}
		
		\def\R{.7}
		
		\pgfmathsetmacro{\RM}{2*\R}
		\pgfmathsetmacro{\Rq}{1.2*\R}
		
		\draw (0,0) ++(40:\R) arc[start angle=40, end angle=-220, radius=\R];
		
		\draw (0:\R) -- (0:\RM);
		\draw (180:\R) -- (180:\RM)node[pos=1, left] {$p_2$};
		\draw (270:\R) -- (270:\RM)node[pos=1, below] {$p_1$};
		
		\draw[->]
		(0,0) ++(-220:\Rq)
		arc[start angle=-220, end angle=-190, radius=\Rq]
		node[midway, left, yshift=4pt] {$q$};
		
		\draw[->]
		(0,0) ++(40:\Rq)
		arc[start angle=40, end angle=10, radius=\Rq]
		node[midway, right, yshift=4pt] {$-q$};
	
	\end{tikzpicture}
	\begin{tikzpicture}
		
		\def\R{.7}
		
		\pgfmathsetmacro{\RM}{2*\R}
		\pgfmathsetmacro{\Rq}{1.2*\R}
		
		\draw (0,0) ++(40:\R) arc[start angle=40, end angle=-220, radius=\R];
		
		\draw (0:\R) -- (0:\RM)node[pos=1, right]{$p_2$};
		\draw (180:\R) -- (180:\RM)node[pos=1, left]{$p_1$} ;
		\draw (270:\R) -- (270:\RM)node[pos=1, below]{$\vphantom{p_1}$} ;
		
		\draw[->]
		(0,0) ++(-220:\Rq)
		arc[start angle=-220, end angle=-190, radius=\Rq]
		node[midway, left, yshift=4pt] {$q$};
		
		\draw[->]
		(0,0) ++(40:\Rq)
		arc[start angle=40, end angle=10, radius=\Rq]
		node[midway, right, yshift=4pt] {$-q$};
	
	\end{tikzpicture}
	\caption{\label{fig:diagramsH3} Independent diagrams contributing to the only channel in $H^{(3)}$, where the continuous internal lines correspond to propagators given in Eq.~\eqref{eq:propagator} and vertices are just $1$.}
\end{figure}
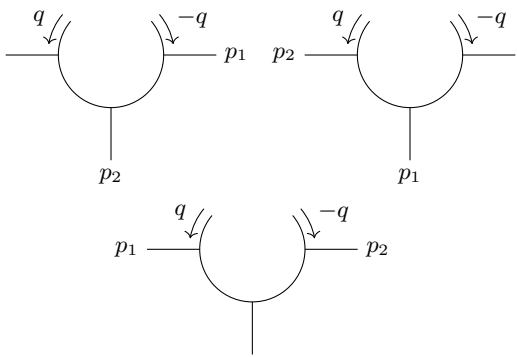

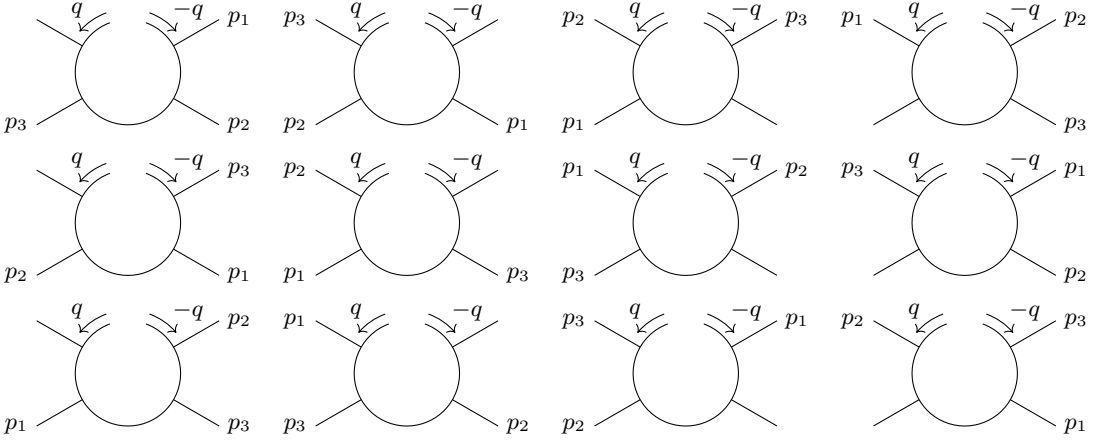
\begin{figure*}[htpb]
	\begin{tikzpicture}
		\def\R{0.7}
		
		\pgfmathsetmacro{\RM}{2*\R}
		\pgfmathsetmacro{\Rq}{1.2*\R}
		
		\draw (0,0) ++(70:\R) arc[start angle=70, end angle=-250, radius=\R];
		
		\draw (30:\R) -- (30:\RM) node[pos=1, right] {$p_1$};
		\draw (-30:\R) -- (-30:\RM)node[pos=1, right] {$p_2$};
		\draw (210:\R) -- (210:\RM)node[pos=1, left] {$p_3$};
		\draw (150:\R) -- (150:\RM)node[pos=1, left] {};

		\draw[->]
		(0,0) ++(110:\Rq)
		arc[start angle=110, end angle=140, radius=\Rq]
		node[midway, left, yshift=4pt] {$q$};
		
		\draw[->]
		(0,0) ++(70:\Rq)
		arc[start angle=70, end angle=40, radius=\Rq]
		node[midway, right, yshift=4pt] {$-q$};
		
	\end{tikzpicture}~
	\begin{tikzpicture}
		\def\R{0.7}
		
		\pgfmathsetmacro{\RM}{2*\R}
		\pgfmathsetmacro{\Rq}{1.2*\R}
		
		\draw (0,0) ++(70:\R) arc[start angle=70, end angle=-250, radius=\R];
		
		\draw (30:\R) -- (30:\RM) node[pos=1, right] {};
		\draw (-30:\R) -- (-30:\RM)node[pos=1, right] {$p_1$};
		\draw (210:\R) -- (210:\RM)node[pos=1, left] {$p_2$};
		\draw (150:\R) -- (150:\RM)node[pos=1, left] {$p_3$};
		
		\draw[->]
		(0,0) ++(110:\Rq)
		arc[start angle=110, end angle=140, radius=\Rq]
		node[midway, left, yshift=4pt] {$q$};
		
		\draw[->]
		(0,0) ++(70:\Rq)
		arc[start angle=70, end angle=40, radius=\Rq]
		node[midway, right, yshift=4pt] {$-q$};
	\end{tikzpicture}~
	\begin{tikzpicture}
		\def\R{0.7}
		
		\pgfmathsetmacro{\RM}{2*\R}
		\pgfmathsetmacro{\Rq}{1.2*\R}
		
		\draw (0,0) ++(70:\R) arc[start angle=70, end angle=-250, radius=\R];
		
		\draw (30:\R) -- (30:\RM) node[pos=1, right] {$p_3$};
		\draw (-30:\R) -- (-30:\RM)node[pos=1, right] {};
		\draw (210:\R) -- (210:\RM)node[pos=1, left] {$p_1$};
		\draw (150:\R) -- (150:\RM)node[pos=1, left] {$p_2$};
		
		\draw[->]
		(0,0) ++(110:\Rq)
		arc[start angle=110, end angle=140, radius=\Rq]
		node[midway, left, yshift=4pt] {$q$};
		
		\draw[->]
		(0,0) ++(70:\Rq)
		arc[start angle=70, end angle=40, radius=\Rq]
		node[midway, right, yshift=4pt] {$-q$};
	\end{tikzpicture}~
	\begin{tikzpicture}
		\def\R{0.7}
		
		\pgfmathsetmacro{\RM}{2*\R}
		\pgfmathsetmacro{\Rq}{1.2*\R}
		
		\draw (0,0) ++(70:\R) arc[start angle=70, end angle=-250, radius=\R];
		
		\draw (30:\R) -- (30:\RM) node[pos=1, right] {$p_2$};
		\draw (-30:\R) -- (-30:\RM)node[pos=1, right] {$p_3$};
		\draw (210:\R) -- (210:\RM)node[pos=1, left] {};
		\draw (150:\R) -- (150:\RM)node[pos=1, left] {$p_1$};
		
		\draw[->]
		(0,0) ++(110:\Rq)
		arc[start angle=110, end angle=140, radius=\Rq]
		node[midway, left, yshift=4pt] {$q$};
		
		\draw[->]
		(0,0) ++(70:\Rq)
		arc[start angle=70, end angle=40, radius=\Rq]
		node[midway, right, yshift=4pt] {$-q$};
	\end{tikzpicture}
	
	\begin{tikzpicture}
		\def\R{0.7}
		
		\pgfmathsetmacro{\RM}{2*\R}
		\pgfmathsetmacro{\Rq}{1.2*\R}
		
		\draw (0,0) ++(70:\R) arc[start angle=70, end angle=-250, radius=\R];
		
		\draw (30:\R) -- (30:\RM) node[pos=1, right] {$p_3$};
		\draw (-30:\R) -- (-30:\RM)node[pos=1, right] {$p_1$};
		\draw (210:\R) -- (210:\RM)node[pos=1, left] {$p_2$};
		\draw (150:\R) -- (150:\RM)node[pos=1, left] {};

		\draw[->]
		(0,0) ++(110:\Rq)
		arc[start angle=110, end angle=140, radius=\Rq]
		node[midway, left, yshift=4pt] {$q$};
		
		\draw[->]
		(0,0) ++(70:\Rq)
		arc[start angle=70, end angle=40, radius=\Rq]
		node[midway, right, yshift=4pt] {$-q$};
		
	\end{tikzpicture}~
	\begin{tikzpicture}
		\def\R{0.7}
		
		\pgfmathsetmacro{\RM}{2*\R}
		\pgfmathsetmacro{\Rq}{1.2*\R}
		
		\draw (0,0) ++(70:\R) arc[start angle=70, end angle=-250, radius=\R];
		
		\draw (30:\R) -- (30:\RM) node[pos=1, right] {};
		\draw (-30:\R) -- (-30:\RM)node[pos=1, right] {$p_3$};
		\draw (210:\R) -- (210:\RM)node[pos=1, left] {$p_1$};
		\draw (150:\R) -- (150:\RM)node[pos=1, left] {$p_2$};
		
		\draw[->]
		(0,0) ++(110:\Rq)
		arc[start angle=110, end angle=140, radius=\Rq]
		node[midway, left, yshift=4pt] {$q$};
		
		\draw[->]
		(0,0) ++(70:\Rq)
		arc[start angle=70, end angle=40, radius=\Rq]
		node[midway, right, yshift=4pt] {$-q$};
	\end{tikzpicture}~
	\begin{tikzpicture}
		\def\R{0.7}
		
		\pgfmathsetmacro{\RM}{2*\R}
		\pgfmathsetmacro{\Rq}{1.2*\R}
		
		\draw (0,0) ++(70:\R) arc[start angle=70, end angle=-250, radius=\R];
		
		\draw (30:\R) -- (30:\RM) node[pos=1, right] {$p_2$};
		\draw (-30:\R) -- (-30:\RM)node[pos=1, right] {};
		\draw (210:\R) -- (210:\RM)node[pos=1, left] {$p_3$};
		\draw (150:\R) -- (150:\RM)node[pos=1, left] {$p_1$};
		
		\draw[->]
		(0,0) ++(110:\Rq)
		arc[start angle=110, end angle=140, radius=\Rq]
		node[midway, left, yshift=4pt] {$q$};
		
		\draw[->]
		(0,0) ++(70:\Rq)
		arc[start angle=70, end angle=40, radius=\Rq]
		node[midway, right, yshift=4pt] {$-q$};
	\end{tikzpicture}~
	\begin{tikzpicture}
		\def\R{0.7}
		
		\pgfmathsetmacro{\RM}{2*\R}
		\pgfmathsetmacro{\Rq}{1.2*\R}
		
		\draw (0,0) ++(70:\R) arc[start angle=70, end angle=-250, radius=\R];
		
		\draw (30:\R) -- (30:\RM) node[pos=1, right] {$p_1$};
		\draw (-30:\R) -- (-30:\RM)node[pos=1, right] {$p_2$};
		\draw (210:\R) -- (210:\RM)node[pos=1, left] {};
		\draw (150:\R) -- (150:\RM)node[pos=1, left] {$p_3$};
		
		\draw[->]
		(0,0) ++(110:\Rq)
		arc[start angle=110, end angle=140, radius=\Rq]
		node[midway, left, yshift=4pt] {$q$};
		
		\draw[->]
		(0,0) ++(70:\Rq)
		arc[start angle=70, end angle=40, radius=\Rq]
		node[midway, right, yshift=4pt] {$-q$};
	\end{tikzpicture}
	
	\begin{tikzpicture}
		\def\R{0.7}
		
		\pgfmathsetmacro{\RM}{2*\R}
		\pgfmathsetmacro{\Rq}{1.2*\R}
		
		\draw (0,0) ++(70:\R) arc[start angle=70, end angle=-250, radius=\R];
		
		\draw (30:\R) -- (30:\RM) node[pos=1, right] {$p_2$};
		\draw (-30:\R) -- (-30:\RM)node[pos=1, right] {$p_3$};
		\draw (210:\R) -- (210:\RM)node[pos=1, left] {$p_1$};
		\draw (150:\R) -- (150:\RM)node[pos=1, left] {};

		\draw[->]
		(0,0) ++(110:\Rq)
		arc[start angle=110, end angle=140, radius=\Rq]
		node[midway, left, yshift=4pt] {$q$};
		
		\draw[->]
		(0,0) ++(70:\Rq)
		arc[start angle=70, end angle=40, radius=\Rq]
		node[midway, right, yshift=4pt] {$-q$};
		
	\end{tikzpicture}~
	\begin{tikzpicture}
		\def\R{0.7}
		
		\pgfmathsetmacro{\RM}{2*\R}
		\pgfmathsetmacro{\Rq}{1.2*\R}
		
		\draw (0,0) ++(70:\R) arc[start angle=70, end angle=-250, radius=\R];
		
		\draw (30:\R) -- (30:\RM) node[pos=1, right] {};
		\draw (-30:\R) -- (-30:\RM)node[pos=1, right] {$p_2$};
		\draw (210:\R) -- (210:\RM)node[pos=1, left] {$p_3$};
		\draw (150:\R) -- (150:\RM)node[pos=1, left] {$p_1$};
		
		\draw[->]
		(0,0) ++(110:\Rq)
		arc[start angle=110, end angle=140, radius=\Rq]
		node[midway, left, yshift=4pt] {$q$};
		
		\draw[->]
		(0,0) ++(70:\Rq)
		arc[start angle=70, end angle=40, radius=\Rq]
		node[midway, right, yshift=4pt] {$-q$};
	\end{tikzpicture}~
	\begin{tikzpicture}
		\def\R{0.7}
		
		\pgfmathsetmacro{\RM}{2*\R}
		\pgfmathsetmacro{\Rq}{1.2*\R}
		
		\draw (0,0) ++(70:\R) arc[start angle=70, end angle=-250, radius=\R];
		
		\draw (30:\R) -- (30:\RM) node[pos=1, right] {$p_1$};
		\draw (-30:\R) -- (-30:\RM)node[pos=1, right] {};
		\draw (210:\R) -- (210:\RM)node[pos=1, left] {$p_2$};
		\draw (150:\R) -- (150:\RM)node[pos=1, left] {$p_3$};
		
		\draw[->]
		(0,0) ++(110:\Rq)
		arc[start angle=110, end angle=140, radius=\Rq]
		node[midway, left, yshift=4pt] {$q$};
		
		\draw[->]
		(0,0) ++(70:\Rq)
		arc[start angle=70, end angle=40, radius=\Rq]
		node[midway, right, yshift=4pt] {$-q$};
	\end{tikzpicture}~
	\begin{tikzpicture}
		\def\R{0.7}
		
		\pgfmathsetmacro{\RM}{2*\R}
		\pgfmathsetmacro{\Rq}{1.2*\R}
		
		\draw (0,0) ++(70:\R) arc[start angle=70, end angle=-250, radius=\R];
		
		\draw (30:\R) -- (30:\RM) node[pos=1, right] {$p_3$};
		\draw (-30:\R) -- (-30:\RM)node[pos=1, right] {$p_1$};
		\draw (210:\R) -- (210:\RM)node[pos=1, left] {};
		\draw (150:\R) -- (150:\RM)node[pos=1, left] {$p_2$};
		
		\draw[->]
		(0,0) ++(110:\Rq)
		arc[start angle=110, end angle=140, radius=\Rq]
		node[midway, left, yshift=4pt] {$q$};
		
		\draw[->]
		(0,0) ++(70:\Rq)
		arc[start angle=70, end angle=40, radius=\Rq]
		node[midway, right, yshift=4pt] {$-q$};
	\end{tikzpicture}
	\caption{\label{fig:diagramsH4} Independent diagrams contributing to $H^{(4)}$, where the continuous internal lines correspond to propagators  given in Eq~.\eqref{eq:propagator} and vertices are just $1$. Notice that in every column, diagrams are equivalent up to a cyclic permutation of the explicit external momenta ($p_1$, $p_2$, $p_3$), while diagrams within a row are equivalent up to a cyclic rotation of \emph{all} external momenta. We thus refer to each row of diagrams as a different ``channel''.}
\end{figure*}

\section{Conformal symmetry of the large-$N$ limit vertex by vertex} \label{sec:conformal}

While the functional approach in Sect.~\ref{sec:functionalProof} elegantly establishes the presence of conformal symmetry, it does not make the internal diagrammatic mechanisms transparent. To understand exactly how the potentially obstructing terms dynamically suppress one another, a necessary step for generalizing to finite $N$, we now explicitly prove conformal invariance vertex by vertex.

To this end, we prove that given the form of the WF fixed point solution of the large-$N$ limit, conformal symmetry is realized exactly at the NPRG fixed point for every vertex function.
Let us first discuss the structure of the equations that we will encounter, and their decomposition into channels. We will take as an example the flow equations of the $n$-point vertex functions $\gamma^{(n)}(p_1,\dots,p_{n-1})$, which can be obtained from Eq.~\eqref{eq:flow_gamma}. They are of the form:
\begin{equation}
    \begin{split}
        &\partial_t\gamma^{(n)}(p_1,\dots,p_{n-1})=\\
        &\frac{(-1)^{n-1}N}{2}\int_{q}\partial_tR_k(q^2)G^{2}(q)H^{(n)}(q,q',p_1,\dots,p_{n-1})\Big|_{q'=-q},
    \end{split}
\end{equation}
where the propagator at uniform field and after a Fourier transformation takes the form:
\begin{equation}\label{eq:propagator}
	G(q)=\frac{1}{q^2+R_k(q^2)+\sigma},
\end{equation}
and the function $H^{(n)}$, which is the Fourier transform of
\begin{equation}
\begin{split}
    H^{(n)}&(x,y,x_1,\dots,x_n)=\\\int_{z,w}&G^{-1}(x,z)\frac{\delta^n G[z,w]}{\delta\sigma(x_1)\dots\delta\sigma(x_n)}G^{-1}(w,y)\Bigg|_{x_n=0},
\end{split}
\end{equation} can be written as the sum over the $(n-1)!/2$ channel contributions
\begin{equation}\label{eq:channelHj}	H^{(n)}=2\sum_{\text{channels}\,j}H^{(n)}_j,
\end{equation}
characterized by the specific cyclic ordering in which the external momenta enter the loops contributions to the flow of $\gamma^{(n)}$. The factor 2 comes from the fact that every diagram is repeated with its internal momentum routing entirely inverted, which remains valid inside the loop integral. Let us stress that we keep the notation $H^{(n)}$ used before in the analogous case of the flow of $\Gamma^{(n)}$ but the $H^{(n)}$ functions for $\gamma^{(n)}$ are simpler. In particular, they do not have indices and all 3-point vertices must be replaced by $1$ while higher order vertices are zero. To be concrete, let us discuss the simpler cases corresponding to vertices $\gamma^{(2)}$, $\gamma^{(3)}$ and $\gamma^{(4)}$. The corresponding diagrams are presented in Figs.~\ref{fig:diagramsH2},~\ref{fig:diagramsH3},~\ref{fig:diagramsH4}. Let us comment that in the case of $\gamma^{(2)}$ there is a single independent external momenta and, consequently, only one channel and diagram. In the case of $\gamma^{(3)}$ there are two independent momenta $p_1$ and $p_2$. Consequently, there are three diagrams. As mentioned before, in principle there would be more diagrams, however, these are equivalent to the previous ones and correspond to the same channel. The full complexity appears from $\gamma^{(4)}$ on. In that case there are 12 diagrams (without inverting order) that are grouped in 3 inequivalent channels $(p_1,p_2,p_3)$, $(p_2,p_3,p_1)$, $(p_3,p_1,p_2)$. They are not cyclically equivalent because as shown in Fig.~\ref{fig:diagramsH4} there is the fourth leg with momenta $-p_1-p_2-p_3$ that is always put in the last position.
In the general case of $\gamma^{(n)}$ with $n\geq 3$, there are $(n-1)!/2$ independent channels with $n$ diagrams for each one (not considering the inverted ordered diagrams).

Let us now turn our attention to conformal WI for the large-$N$ limit of $O(N)$ models, in terms of the functions $\gamma^{(n)}(p_1,\dots,p_{n-1})$. In particular, we are interested in the identities related to rotations, dilations and special conformal transformations, which take the following forms:
\begin{widetext}
	\textit{Rotations}
	\begin{equation}\label{eq:RotEq}
		\mathcal{J}_{\mu\nu}\gamma^{(n)}(p_1,\dots,p_{n-1})\equiv\sum_{i=1}^{n-1}\big(p_{i\mu}\partial_{i\nu}-p_{i\nu}\partial_{i\mu}\big)\gamma^{(n)}(p_1,\dots,p_{n-1})=0,
	\end{equation}
	\textit{Dilations:}
	\begin{equation}\label{eq:DilEq}
    \begin{split}
			\mathcal{D}\gamma^{(n)}(p_1,\dots,p_{n-1})\equiv\Big(-\bigg[\sum_{i=1}^{n-1} p_{i\nu}\partial_{i\nu}\bigg]+(2D_{\varphi}-d)n&+d+(2D_{\varphi}-d)\sigma\partial_\sigma\Big)\gamma^{(n)}(p_1,\dots,p_{n-1})
			=\\&\frac{(-1)^{n-1}N}{2}\int_q \partial_t R_k(q^2)G^2(q) H^{(n)}(q,-q,p_1,\dots,p_{n-1}),
    \end{split}
	\end{equation}
	\textit{Special conformal transformations (SCT):}
	\begin{equation}
		\begin{split}
			\mathcal{K}_\mu\gamma^{(n)}(p_1,\dots,p_{n-1})\equiv\Big(\sum_{i=1}^{n-1}\big[p_{i\mu}\partial_{i\nu}\partial_{i\nu}&-2p_{i\nu}\partial_{i\nu}\partial_{i\mu}+2(2D_{\varphi}-d)\partial_{i\mu}\big]\Big)\gamma^{(n)}(p_1,\dots,p_{n-1})\\&+2(2D_{\varphi}-d)\sigma\partial_{r_{\mu}}\gamma^{(n+1)}(r,p_1,\dots,p_{n-1})\Big|_{r=0}=
			\\\frac{(-1)^{n-1}N}{2}&\int_q\partial_t R_k(q^2)G^2(q)\big(\partial_{q_{\mu}}+\partial_{q'_{\mu}}\big) H^{(n)}(q,q'p_1,\dots,p_{n-1})\Big|_{q'=-q}.
		\end{split}
	\end{equation}
\end{widetext}

These expressions are obtained by differentiating the WIs (given in \eqref{eq:dilgammaapp} for dilatations and \eqref{eq:confgammaapp} for SCT) for the different symmetries $n$ times with respect to $\sigma$, then evaluating them in a uniform field, and finally performing a Fourier transform. The translation identity is not explicitly stated, as it follows implicitly from the requirement that total momentum be conserved.

It is interesting to note that equation (\ref{eq:DilEq}) is identical to the fixed-point equation for the flux under the NPRG of the {\it dimensionless} $\gamma^{(n)}$. This is a general property: the WI for dilations is always identical to the fixed-point condition of the NPRG in any model. For a detailed discussion of this point, see \cite{delamotte2016scale,Dupuis:2020fhh}.

The proof we provide here regarding the realization of conformal symmetry in the large-$N$ limit, works for every channel independently, i.e. each contribution cancels by itself for each identity. This is obvious for rotations (each channel in $\gamma^{(n)}$ is scalar, so rotationally invariant). In the case of dilatation invariance, the result is due to the fact that
\begin{equation}
\begin{split}
	\int_q \partial_t R_k(q^2)G^2(q)&H^{(n)}_{j}\big(q,-q,p_{P_j(1)},\dots,p_{P_j(n-1)}\big)\\&=-\partial_t I^{(n)}\big(p_{P_j(1)},\dots,p_{P_j(n-1)}\big),
\end{split}
\end{equation}
for the permutation $P_j\{i\}$ of the explicit external momenta (i.e. $i\in \lbrace 1,\dots,n-1\rbrace$) yielding the independent channel $j$, that is shown in Appendix~\ref{app:indepChanDil}. 
For SCT this is far from trivial and is one of the results of this work. That is, at large-$N$, channels can be readily identified and are decoupled when considering the functional $\gamma[\sigma]$. This, perhaps, is just a peculiarity of the large-$N$ limit. However, it may correspond to a more general structure valid also at finite $N$. In fact, in the present case, it is enough to prove that SCT WI is verified for any particular channel.

It proves useful to introduce the following definitions of integrals similar to the $I^{(n)}$'s that we summarize together with the definition of $I^{(n)}$: 
\begin{widetext}

    \begin{equation}
		I^{(n)}(p_1,\dots,p_{n-1})\equiv\int_q \prod_{j=0}^{n-1}G(Q_j),
	\end{equation}

    \begin{equation}\label{eq:Inp1idef}
    \begin{split}
        I^{(n+1)}_{i}(r;p_1,\dots,p_{n-1})\equiv &I^{(n+1)}(p_1,\dots,p_i,r,p_{i+1},\dots,p_{n-1})\\&=\int_q G(q)G(q+r)\Big\lbrace\prod_{j=1}^{n-i-1}G\Big(q+r+\sum_{l=1}^{j}p_{i+l}\Big)\Big\rbrace\Big\lbrace\prod_{j=1}^{i}G\Big(q'+\sum_{l=1}^{j}p_{i-l+1}\Big)\Big\rbrace_{q'=-q},
    \end{split}
	\end{equation}
    \begin{equation}\label{eq:Jndef}
		J^{(n)}_{i}(p_1,\dots,p_{n-1})\equiv\int_q \partial_tR_k(q^2)G(q)^2\Big\lbrace\prod_{j=1}^{n-i-1}G\Big(q+\sum_{l=1}^{j}p_{i+l}\Big)\Big\rbrace\Big\lbrace\prod_{j=1}^{i}G\Big(q'+\sum_{l=1}^{j}p_{i-l+1}\Big)\Big\rbrace_{q'=-q},
	\end{equation}
	\begin{equation}\label{eq:Jmundef}
		J^{(n)}_{i,\mu}(p_1,\dots,p_{n-1})\equiv\int_q \partial_tR_k(q^2)G^2(q)\big(\partial _{q_{\mu}}+\partial _{q'_{\mu}}\big)\Big\lbrace\prod_{j=1}^{n-i-1}G\Big(q+\sum_{l=1}^{j}p_{i+l}\Big)\Big\rbrace\Big\lbrace\prod_{j=1}^{i}G\Big(q'+\sum_{l=1}^{j}p_{i-l+1}\Big)\Big\rbrace_{q'=-q},
	\end{equation}
\end{widetext}
with the index $i$ ranging between 0 and $n-1$, $Q_j=q+\sum_{i=1}^j p_i$ and $Q_0=q$.

\begin{figure}[ht]
	\centering
	\begin{tikzpicture}
		\begin{feynman}
			
			\def\R{.7}
			
			\pgfmathsetmacro{\RM}{2*\R}
			\pgfmathsetmacro{\rc}{0.2*\R}
			\pgfmathsetmacro{\Rq}{1.2*\R}
			\pgfmathsetmacro{\Rp}{1.4*\R}
			
			\vertex (a) at (0,0);
			
			\vertex (b) at (90:\R);
			
			\vertex (c) at (170:\Rp);
			\vertex (d) at (180:\Rp);
			\vertex (e) at (190:\Rp);

			\draw (0,0) circle (\R);
			
			\draw (30:\R) -- (30:\RM) node[pos=1, right] {$p_1$};
			\draw (-30:\R) -- (-30:\RM)node[pos=1, right] {$p_2$};
			\draw (-90:\R) -- (-90:\RM)node[pos=1, below] {$p_3$};
			\draw (210:\R) -- (210:\RM)node[pos=1, left] {$p_4$};
			\draw (150:\R) -- (150:\RM)node[pos=1, left] {$p_n$};
			
			\draw (b) ++(-\rc,-\rc) -- ++(2*\rc,2*\rc)
			++(-2*\rc,0) -- ++(2*\rc,-2*\rc);
			
			\draw[->]
			(0,0) ++(105:\Rq)
			arc[start angle=105, end angle=140, radius=\Rq]
			node[midway, left, yshift=4pt] {$q$};
			
			\draw[->]
			(0,0) ++(75:\Rq)
			arc[start angle=75, end angle=40, radius=\Rq]
			node[midway, right, yshift=4pt] {$-q$};
			
			\filldraw[black] (c) circle (0.5pt);
			\filldraw[black] (d) circle (0.5pt);
			\filldraw[black] (e) circle (0.5pt);
			
			\node at (3.1*\R,0) {$+\;\dots\;+$};
			
		\end{feynman}
	\end{tikzpicture}
	\begin{tikzpicture}
		\begin{feynman}
				
			\def\R{.7}
			
			\pgfmathsetmacro{\RM}{2*\R}
			\pgfmathsetmacro{\rc}{0.2*\R}
			\pgfmathsetmacro{\Rq}{1.2*\R}
			\pgfmathsetmacro{\Rp}{1.4*\R}
			
			\vertex (a) at (0,0);
			
			\vertex (b) at (90:\R);
			
			\vertex (c) at (170:\Rp);
			\vertex (d) at (180:\Rp);
			\vertex (e) at (190:\Rp);

			\draw (0,0) circle (\R);
			
			\draw (30:\R) -- (30:\RM) node[pos=1, right] {$p_n$};
			\draw (-30:\R) -- (-30:\RM)node[pos=1, right] {$p_1$};
			\draw (-90:\R) -- (-90:\RM)node[pos=1, below] {$p_2$};
			\draw (210:\R) -- (210:\RM)node[pos=1, left] {$p_3$};
			\draw (150:\R) -- (150:\RM)node[pos=1, left] {$p_{n-1}$};
			
			\draw (b) ++(-\rc,-\rc) -- ++(2*\rc,2*\rc)
			++(-2*\rc,0) -- ++(2*\rc,-2*\rc);
			
			\draw[->]
			(0,0) ++(105:\Rq)
			arc[start angle=105, end angle=140, radius=\Rq]
			node[midway, left, yshift=4pt] {$q$};
			
			\draw[->]
			(0,0) ++(75:\Rq)
			arc[start angle=75, end angle=40, radius=\Rq]
			node[midway, right, yshift=4pt] {$-q$};
			
			\filldraw[black] (c) circle (0.5pt);
			\filldraw[black] (d) circle (0.5pt);
			\filldraw[black] (e) circle (0.5pt);
			
		\end{feynman}
	\end{tikzpicture}
	\caption{Diagrammatic channel contribution to the flow of $\gamma^{(n)}$ corresponding to channel 1. The cross on top stands for $\partial_t R_k$, internal lines represent the propagator $G(q)$ given in Eq.~\eqref{eq:propagator}, while the loop represents the integration over the internal momenta $\int_q$. Vertices are just $1$ and external legs are amputated.}\label{fig:channelContrFlow}
\end{figure}
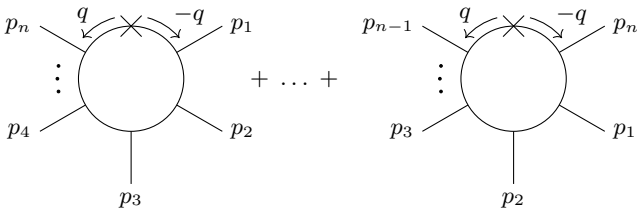

As explained before, it is enough to prove the SCT identity in a particular channel. Without loss of generality, we will consider the simple channel $p_1,p_2,p_3,\cdots,p_{n-1}$ and omit the explicit momentum dependence of functions to simplify notation. We will refer to it as channel 1, and it is depicted in Fig.~\ref{fig:channelContrFlow}.

In the SCT equation for $\gamma^{(n)}$, there is a term where the vertex $\gamma^{(n+1)}$ with an extra momentum, denoted here as $r$, appears. 
It should be noted that for a given value of $n$, the number of permutations of the external momenta in $\gamma^{(n+1)}$ is $n$ times the number of permutations in $\gamma^{(n)}$. This is because in every channel there are $n$ possible places in which the extra momentum $r$ can enter the loop. Continuing with this idea, $\gamma^{(n+1)}$ is just the sum of the $n$ insertions of $r^{\mu}$ in each of the $(n-1)!$ channels -- with a global $(-1)$ factor with respect to $\gamma^{(n)}$. For a given channel in the function $I^{(n)}$ we denote the $n$ insertions of the momentum $r$ as $I^{(n+1)}_j$ with $j$ between 0 and $n-1$, see Eq.~\eqref{eq:Inp1idef}.

In terms of the functions $I^{(n)}$'s, $I^{(n)}_j$'s, $J^{(n)}_j$'s and $J^{(n)}_{j,\mu}$'s the modified WI for each channel read:
\begin{widetext}
	\textit{Rotations in a channel}
	\begin{equation}\label{eq:ChanRotEq}
    \mathcal{J}_{\mu\nu}I^{(n)}\equiv\sum_{i=1}^{n-1}\big(p_{i\mu}\partial_{i\nu}-p_{i\nu}\partial_{i\mu}\big)I^{(n)}=0,
	\end{equation}
	\textit{Dilations in a channel:}
	\begin{equation}\label{eq:ChanDilEq}
		\mathcal{D}I^{(n)}\equiv\Big(-\bigg[\sum_{i=1}^{n-1} p_{i\nu}\partial_{i\nu}\bigg]+(2D_{\varphi}-d)n+d+(2D_{\varphi}-d)\sigma\partial_\sigma\Big)I^{(n)}
		=-\sum_{j=0}^{n-1}J^{(n)}_j,
	\end{equation}
	\textit{Special conformal transformations in a channel:}
	\begin{equation}\label{eq:ChanSCTEq}
		\mathcal{K}_\mu I^{(n)}\equiv\Big(\sum_{i=1}^{n-1}\big[p_{i\mu}\partial_{i\nu}\partial_{i\nu}-2p_{i\nu}\partial_{i\nu}\partial_{i\mu}+2(2D_{\varphi}-d)\partial_{i\mu}\big]\Big)I^{(n)}-2(2D_{\varphi}-d)\sigma\partial_{r_{\mu}}\sum_{j=0}^{n-1}I^{(n+1)}_j\Big|_{r=0}\stackrel{?}{=}-\sum_{j=0}^{n-1}J^{(n)}_{j,\mu}.
	\end{equation}
\end{widetext}
The first two identities will be assumed to be satisfied (that is, the system is rotationally invariant and scale/dilationally invariant), and the goal is to prove the third one. Proving that the first two imply the last is equivalent to proving that scale invariance (plus rotation and translation invariance) implies conformal invariance.

At first glance, the most problematic terms in the SCT identity are: in the left-hand side, the momentum derivative with respect to the momentum $r_{\mu}$ which is then set to zero and the derivatives within the loop integral on the right-hand side. We begin our analysis with the former. We show in App.~\ref{app:drmuandqmuH} that the derivative with respect to $r_{\mu}$ for a channel yields:
\begin{equation}\label{eq:drmu}
	\begin{split}
		-2\partial_{r_{\mu}}\sum_{j=0}^{n-1} I^{(n+1)}_j&\Big|_{r=0} =\bigg(\sum_{i=1}^{n-1}\partial_{i\mu}\bigg)\partial_{\sigma}I^{(n)}\\&+
		\sum_{i=1}^{n-2}\sum_{j=i+1}^{n-1}\Big[\partial_{i\mu}I^{(n+1)}_j+\partial_{j\mu}I^{(n+1)}_{i-1}\Big]\Big|_{r=0}.
	\end{split}
\end{equation}
Notice that for the case $n=2$ the cross terms appearing in the second line of Eq.~\eqref{eq:drmu} do not exist. Therefore, this property is equivalent to
\begin{equation}\label{eq:drmuGamma3}
	\partial_{r_\mu}\Gamma_{iab}^{(3)}(r,p)\Big|_{r=0}=\frac{1}{2}\partial_{p_\mu}\Gamma_{iab}^{(3)}(0,p),
\end{equation}
used in \cite{Cabrera2025}, which is one of the key ingredients to show that for $\Gamma^{(2)}$, the modified dilation WI implies automatically also the modified SCT WI for arbitrary $N$. 

\begin{figure}[t!]
	\centering
	
	\begin{tikzpicture}
		\begin{feynman}
			
			\def\R{.7}
			
			\pgfmathsetmacro{\RM}{2*\R}
			\pgfmathsetmacro{\rc}{0.2*\R}
			\pgfmathsetmacro{\Rq}{1.2*\R}
			\pgfmathsetmacro{\Rp}{1.4*\R}
			
			\vertex (a) at (0,0);
			
			\vertex (b) at (90:\R);
			
			\vertex (c) at (170:\Rp);
			\vertex (d) at (180:\Rp);
			\vertex (e) at (190:\Rp);
			
			\draw (0,0) circle (\R);
			
			\draw (30:\R) -- (30:\RM) node[pos=1, right] {$p_1$};
			\draw (-30:\R) -- (-30:\RM)node[pos=1, right] {$p_2$};
			\draw (-90:\R) -- (-90:\RM)node[pos=1, below] {$p_3$};
			\draw (210:\R) -- (210:\RM)node[pos=1, left] {$p_4$};
			\draw (150:\R) -- (150:\RM)node[pos=1, left] {$p_n$};
			
			\draw (b) ++(-\rc,-\rc) -- ++(2*\rc,2*\rc)
			++(-2*\rc,0) -- ++(2*\rc,-2*\rc);
			
			\draw[thick] (120:0.92*\R) -- (120:1.08*\R);
			\draw[thick] (60:0.92*\R) -- (60:1.08*\R);

			\draw[->]
			(0,0) ++(105:\Rq)
			arc[start angle=105, end angle=140, radius=\Rq]
			node[midway, left, yshift=4pt] {$q$};
			
			\draw[->]
			(0,0) ++(75:\Rq)
			arc[start angle=75, end angle=40, radius=\Rq]
			node[midway, right, yshift=4pt] {$q'$};
			
			\filldraw[black] (c) circle (0.5pt);
			\filldraw[black] (d) circle (0.5pt);
			\filldraw[black] (e) circle (0.5pt);
			\node at (3.1*\R,0) {$+\;\dots\;+$};
		\end{feynman}
	\end{tikzpicture}
	\begin{tikzpicture}
		\begin{feynman}
			
			\def\R{.7}
			
			\pgfmathsetmacro{\RM}{2*\R}
			\pgfmathsetmacro{\rc}{0.2*\R}
			\pgfmathsetmacro{\Rq}{1.2*\R}
			\pgfmathsetmacro{\Rp}{1.4*\R}
			
			\vertex (a) at (0,0);
			
			\vertex (b) at (90:\R);
			
			\vertex (c) at (170:\Rp);
			\vertex (d) at (180:\Rp);
			\vertex (e) at (190:\Rp);

			\draw (0,0) circle (\R);
			
			\draw (30:\R) -- (30:\RM) node[pos=1, right] {$p_n$};
			\draw (-30:\R) -- (-30:\RM)node[pos=1, right] {$p_1$};
			\draw (-90:\R) -- (-90:\RM)node[pos=1, below] {$p_2$};
			\draw (210:\R) -- (210:\RM)node[pos=1, left] {$p_3$};
			\draw (150:\R) -- (150:\RM)node[pos=1, left] {$p_{n-1}$};
			
			\draw (b) ++(-\rc,-\rc) -- ++(2*\rc,2*\rc)
			++(-2*\rc,0) -- ++(2*\rc,-2*\rc);
			
			\draw[thick] (120:0.92*\R) -- (120:1.08*\R);
			\draw[thick] (60:0.92*\R) -- (60:1.08*\R);
			
			\draw[->]
			(0,0) ++(105:\Rq)
			arc[start angle=105, end angle=140, radius=\Rq]
			node[midway, left, yshift=4pt] {$q$};
			
			\draw[->]
			(0,0) ++(75:\Rq)
			arc[start angle=75, end angle=40, radius=\Rq]
			node[midway, right, yshift=4pt] {$q'$};
			
			\filldraw[black] (c) circle (0.5pt);
			\filldraw[black] (d) circle (0.5pt);
			\filldraw[black] (e) circle (0.5pt);
			
		\end{feynman}
	\end{tikzpicture}
	\caption{Diagrammatic channel contribution to the loop term in the SCT of $\gamma^{(n)}$. The cross on top stands for $\partial_t R$, internal lines represent the propagator $G(q)$ given in Eq.~\eqref{eq:propagator}, the short radial lines represent the derivatives with respect to the $q$ and $q'$ momenta on the lower part of the diagrams and subsequent evaluation at $q'=-q$, while the loop represents the integration over the internal momenta $\int_q$ to be performed at last. The external leg is $p_n=-(p_1+\dots+p_{n-1}+q+q')$ before evaluating $q'=-q$. Vertices are just $1$.}	\label{fig:channelContrConf}
\end{figure}
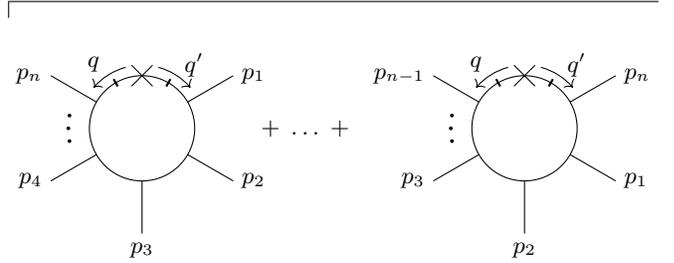

Now, for the case of the loop term with $q$ and $q'$ derivatives, it can be readily seen that a similar structure as the one we have just faced arises. Again, to show this, we can isolate $H_{j=1}^{(n)}$ from $H^{(n)}$, that is, only the $n$ diagrams where the permutation $p_1,\dots,p_{n-1}$ appears in that order. These are the $n$ terms $J^{(n)}_{j,\mu}$ represented as diagrams in Fig.~\ref{fig:channelContrConf}.

We also show in App.~\ref{app:drmuandqmuH} that the loop contribution in the channel 1 is given by:
\begin{equation}\label{eq:dqdqpH}
	\begin{split}
		-\sum_{j=0}^{n-1}J^{(n)}_{j,\mu}&=\Big(\sum_{i=1}^{n-1}\partial_{i\mu}\Big)\partial_{t}I^{(n)}\\
		&+\sum_{i=1}^{n-2}\sum_{j=i+1}^{n-1}\Big[\partial_{i\mu}J_j^{(n)}+\partial_{j\mu}J_{i-1}^{(n)}\Big],
	\end{split}
\end{equation}
where it was observed that, in the large-$N$ limit, we can identify:
\begin{equation}
	\partial_t G(q)\vert_{q,\sigma}=-\partial_t R(q)G^2(q).
\end{equation}
Notice, once again, that for the special case of $n=2$ the cross terms appearing in the second line of Eq.~\eqref{eq:dqdqpH} do not exist which becomes equivalent to the statement: 
\begin{equation}\label{eq:dqdqpmuH2}
    \begin{split}
	   \int_q\partial_t R(q^2)G^2(q)&\big(\partial_{q_{\mu}}+\partial_{q'_{\mu}}\big) H^{(2)}(q,p,q')\Big|_{q'=-q}=\\&\partial_{p\mu}\int_q\partial_t R(q^2)G^2(q) H^{(2)}(q,p,-q),
    \end{split}
\end{equation}
which is the remaining key ingredient to prove the equivalence of dilation and SCT WI used in \cite{Cabrera2025}.

Replacing these contributions into the SCT WI Eq.~\eqref{eq:ChanSCTEq} that we aim to prove and rearranging terms yields:
\begin{widetext}
	\begin{equation}\label{eq:replacedChanSCT}
		\begin{split}
			\sum_{i=1}^{n-1}\big[ p_{i\mu}\partial_{i\nu}\partial_{i\nu}&-2p_{i\nu}\partial_{i\nu}\partial_{i\mu}
			+(2D_{\varphi}-d)\sigma\partial_{\sigma}\partial_{i\mu}+2(2D_{\varphi}-d)\partial_{i\mu}\big]I^{(n)}
			-
			\sum_{i=1}^{n-1}\partial_{i\mu}\partial_{t}I^{(n)}\\&
			\stackrel{?}{=}\sum_{i=1}^{n-2}\sum_{j=i+1}^{n-1}
			\Big[
			\partial_{i\mu} \big(J^{(n)}_j+(d-2D_{\varphi})\sigma I^{(n+1)}_j\Big|_{r=0}\big)
			+\partial_{j\mu}\big(J^{(n)}_{i-1}+(d-2D_{\varphi})\sigma I^{(n+1)}_{i-1}\Big|_{r=0}\big)
			\Big].
		\end{split}
	\end{equation}
\end{widetext}

At this point, we notice that many terms look similar to those of dilation WI, with the difference that there is an extra operator $\big(\sum_{i=1}^{n-1}\partial_{i\mu}\big)$. It will prove useful to consider the combination $\big(\sum_{i=1}^{n-1}\partial_{i\nu}\big)(\delta_{\mu\nu}\mathcal{D}+\mathcal{J}_{\mu\nu})$. Using the expression for rotation and dilation WI given in Eq.~\eqref{eq:ChanRotEq} and Eq.~\eqref{eq:ChanDilEq}, respectively, a straightforward calculation shows that, up to a rearrangement of terms, this is:
\begin{widetext}
\begin{equation}\label{eq:combinedDilRot}
    \begin{split}
        & \sum_{i=1}^{n-1}\big[ p_{i\mu}\partial_{i\nu}\partial_{i\nu}-2p_{i\nu}\partial_{i\nu}\partial_{i\mu}
			+(2D_{\varphi}-d)\sigma\partial_{\sigma}\partial_{i\mu}+n(2D_{\varphi}-d)\partial_{i\mu}\big]I^{(n)}
			-
			\sum_{i=1}^{n-1}\partial_{i\mu}\partial_{t}I^{(n)}=\\&\sum_{i=1}^{n-2}\sum_{j=i+1}^{n-1}\Bigg[
			p_{i\nu} \partial_{j\mu} \partial_{i\nu}+p_{j\nu} \partial_{i\mu} \partial_{j\nu}+p_{i\nu}\partial_{j\nu} \partial_{i\mu}+p_{j\nu}\partial_{i\nu} \partial_{j\mu}-p_{i\mu}\partial_{j\nu} \partial_{i\nu}
			-p_{j\mu}\partial_{i\nu} \partial_{j\nu}
			\Bigg] I^{(n)},
    \end{split}
\end{equation}
\end{widetext}
where it was used, as shown in App.~\ref{app:indepChanDil}, that
\begin{equation}
    \sum_{j=0}^{n-1}J^{(n)}_j=-\partial_t I^{(n)}.
\end{equation}

Equation \eqref{eq:combinedDilRot} is to be compared to Eq.~\eqref{eq:replacedChanSCT}. Observing these expressions we notice the resemblance with the SCT for the channel 1. Indeed, subtracting these two expressions (i.e. performing the operation $\big[\big(\sum_{i=1}^{n-1}\partial_{i\nu}\big)(\delta_{\mu\nu}\mathcal{D}+\mathcal{J}_{\mu\nu})-\mathcal{K}_\mu\big]I^{(n)}$) and rearranging terms yields:
\begin{widetext}
	\begin{equation}\label{eq:combCDJ}
		\begin{split}
			\big[(n-2)&(2D_\varphi-d)\big]
			\sum_{i=1}^{n-1}\partial_{i\mu} I^{(n)}
			\stackrel{?}{=}\\&\sum_{i=1}^{n-2}\sum_{j=i+1}^{n-1}\Bigg[
			p_{i\nu} \partial_{j\mu} \partial_{i\nu}+p_{j\nu} \partial_{i\mu} \partial_{j\nu}+p_{i\nu}\partial_{j\nu} \partial_{i\mu}+p_{j\nu}\partial_{i\nu} \partial_{j\mu}-p_{i\mu}\partial_{j\nu} \partial_{i\nu}
			-p_{j\mu}\partial_{i\nu} \partial_{j\nu}
			\Bigg] I^{(n)}\\
			&-
			\sum_{i=1}^{n-2}\sum_{j=i+1}^{n-1}
			\Big[
			\partial_{i\mu}\big(J^{(n)}_j+(d-2D_\varphi)\sigma I^{(n+1)}_j\Big|_{r=0}\big)
			+\partial_{j\mu}\big(J^{(n)}_{i-1}+(d-2D_\varphi)\sigma I^{(n+1)}_{i-1}\Big|_{r=0}\big)
			\Big].
		\end{split}
	\end{equation}
\end{widetext}
For the remainder of the proof, performed in App.~\ref{app:finalPart}, the last piece of the puzzle (apart from a kinematic algebraic manipulation) is the fact that the fixed point equation for the propagator reads
\begin{equation}\label{eq:propFP}
	\Big(\partial_t+(d-2D_\varphi)\sigma\partial_\sigma+q_\nu\partial_{q_\nu}+2\Big)G(q)=0,
\end{equation}
which is valid only for vanishing anomalous dimension $\eta$.
Using this, it can be shown that Eq.~\eqref{eq:combCDJ} holds exactly and, consequently, the modified special conformal WI for the channel 1 of all vertex functions in the presence of a regulator is indeed satisfied at the fixed point of the large-$N$ limit of $O(N)$ models. Of course, the proof for all other channels is identical and, consequently, it is also valid for the full vertices.

\section{Summary and outlook} \label{sec:discussion}

In this work, we have explicitly proven that in the limit $N\to \infty$ of $O(N)$ models, the critical point is not only invariant under isometries and dilations but also invariant under the entire conformal group. We have proven this in the presence of a {\it non-zero} infrared regulator.  To this end, we focused on the particular analytic solution that is critical (and not multicritical), the Wilson-Fisher fixed point. We have proven that this solution is automatically conformally invariant for $O(N)$ models when $N$ tends to infinity. We have first given a formal functional proof, which has the advantage of being rather short and straightforward. We have then given another proof that analyses, for each vertex of the theory, the behaviour of WI in the presence of the regulator. This is important because our purpose is to understand how the structure of identities is organized in order to clarify how this can happen also for finite values of $N$. In this regard, our work differs from the recent one by Pagani and Sonoda \cite{Pagani:2025dtc}, where a proof of conformal invariance is carried out for the limit $N\to\infty$ of the $O(N)$ models through the analysis of the stress-tensor, and without analyzing the behavior of WI for the vertices. As noted in the introduction, the analysis of WI could easily have been included in that paper. The main contribution of the present paper is the detailed analysis of how the  compensations between diagrams result in something that is far from trivial.

There is a technical point that should be noted that greatly simplifies the proof, which is that the behaviour of the generating functional of the vertices is much simpler for $N\to \infty$ once a certain Legendre transform has been performed \cite{Sonoda:2023ohb}. This points to an important difficulty when extending the present result to finite values of $N$ because, even for $N\to\infty$, the direct proof for vertices defined in the usual way is remarkably complex. Just to give one example, the proof for vertices with three external legs is very simple in the present work, while the direct proof without using the Legendre transform is very cumbersome, see \cite{Cabrera2024}. Another important technical point is that both proofs explicitly use the fact that the anomalous dimension vanishes in the large $N$ limit.

A positive aspect of this work is that we have proven that conformal invariance holds, not only for vertices, but also for certain substructures with particular kinematics (which we have referred to as ``channels'' in the main text). In the case of the limit $N\to\infty$, these particular structures are easy to identify, but their extension to finite $N$ or to theories more general than $O(N)$ models is not entirely obvious (although this work provides possible clues). 

One technical aspect that may be worth mentioning is that a certain combination of identities associated with special conformal transformations, dilations, and rotations seems to play a particular role. This corresponds to the combination
$\big[\big(\sum_{i=1}^{n-1}\partial_{i\nu}\big)(\delta_{\mu\nu}\mathcal{D}+\mathcal{J}_{\mu\nu})-\mathcal{K}_\mu\big]$. This was already pointed out previously when exploring the violation of conformal symmetry due to approximations \cite{Balog2020}. 

This work opens up several possible extensions and perspectives. First, it would be useful to extend the present analysis to finite $N$. To do so, it would probably be interesting to address the expansion in $1/N$ first. This is by no means trivial, since the exact treatment of the $1/N$ expansion has been elusive in the context of NPRG. 

A second natural extension is to analyse conformal constraints not only for the fundamental field but also for correlation functions involving composite operators. This includes scalar operators and, more generally, tensorial ones. In particular, it would be interesting to combine the present work with the study of critical dimensions of vector operators. Previous work has shown that the spectrum of critical dimensions of these operators determines the validity of conformal invariance \cite{Polchinski1988,delamotte2016scale,DePolsi2019}. Then, it would be interesting to use the present analysis to determine these critical dimensions providing indirect proof for the realization of conformal symmetry at any order of the $1/N$ expansion.

Finally, the purpose of employing conformal invariance is to try to improve our understanding of critical phenomena, either exactly or approximately. In recent years \cite{Balog2020,Delamotte2024,Cabrera2025}, we have taken advantage of the constraints arising from conformal invariance with views to improving the quality of approximate schemes such as the Derivative Expansion of the NPRG. In the large $N$ limit, such approximation schemes become, in a sense, exact, and it would be interesting to understand how these two approaches can be combined.

\begin{acknowledgments}
We thank B. Delamotte and M. Tissier for discussions.
The authors acknowledge support from the Programa de Desarrollo de las Ciencias Básicas (PEDECIBA) and the French-Uruguayan Institute of Physics Project (IFUP). GDP and SC acknowledge support from the grant number FCE-3-2024-1-180709 of the Agencia Nacional de Investigación e Innovación (Uruguay). SC acknowledges the support of Comisión Académica de Posgrado de la Universidad de la República. AR has benefited from the financial support of the Grant No. ANR-24-CE30-6695 FUSIoN, and acknowledges the support of the CDP C2EMPI, as well as the French State under the France-2030 programme, the University of Lille, the Initiative of Excellence of the University of Lille, the European Metropolis of Lille for their funding and support of the R-CDP-24-004-C2EMPI project. 
\end{acknowledgments} 

\appendix

\section{Derivation of the modified WI of $\gamma[\sigma]$}
\label{app:WI}

In this Appendix, we derive the modified WI of the functional $\gamma$ from first principles. The modified WI for $\Gamma[\vec\phi]$ are given in Eq.~\eqref{eq:genericWI} for an arbitrary symmetry. Using that in large $N$, 
$	\Gamma[\vec{\phi}]=\int_x\frac{(\nabla\vec{\phi})^2}{2}+\hat{\Gamma}[\rho]$, and that the first term in the right-hand side is invariant under all conformal transformations, the modified WI for $\hat{\Gamma}$ read
\begin{equation}
	\begin{split}
		\int_x \frac{\delta\hat\Gamma}{\delta\rho(x)}\updelta\rho(x)=-\int_{x,y}\frac{\delta\hat\Gamma}{\delta R_{k}(x,y)}\updelta R_{k}(x,y),
	\end{split}
\end{equation}
where $\updelta\rho(x)$ takes a form similar to Eq.~\eqref{eq:TransfField} with $D_\varphi\to D_\rho=2D_\varphi$. We also have specialized to the case where the regulator is explicitly $O(N)$ invariant. The functionals $\hat{\Gamma}[\rho]$ and $\gamma[\sigma]$ are related by a Legendre transformation,
\begin{equation}
    \hat{\Gamma}[\rho]=\int_x\sigma(x)\rho(x)-\gamma[\sigma],
\end{equation}
with $\delta\gamma[\sigma]/\delta\sigma(x)=\rho(x)$, and thus $\delta \hat{\Gamma}[\rho]/\delta R_k(x,y)=-\delta\gamma[\sigma]/\delta R_k(x,y)$. 

By integration by part, one shows that $\int_x \sigma(x)\updelta\rho(x) = -\int_x \updelta\sigma(x)\rho(x)$ with $\updelta\sigma(x)$ taking a form similar to Eq.~\eqref{eq:TransfField} with $D_\varphi\to D_\sigma=d-2D_\varphi$.
This leads to the generic modified WI
\begin{equation}
    -\int_x \frac{\delta\gamma[\sigma]}{\delta\sigma(x)}\updelta\sigma(x)=\int_{x,y}\frac{\delta\gamma[\sigma]}{\delta R_k(x,y)}\updelta R_k(x,y).
\end{equation}
Using $\frac{\delta\gamma[\sigma]}{\delta R_k(x,y)}=-\frac N2 G[x,y]$, one obtains for the dilatation WI (using $\updelta_{\rm dil} R_k=\partial_t R_k$)
\begin{equation}
    \int_x \frac{\delta\gamma[\sigma]}{\delta\sigma(x)}(D^x+d-2D_\varphi)\sigma(x)=\frac N2\int_{x,y}G[x,y]\partial_t R_k(x,y).
    \label{eq:dilgammaapp}
\end{equation}
Differentiating with respect to $\sigma(x)$ and using $\frac{\delta G[x,y]}{\delta \sigma[z]}=-G[x,z]G[z,y]$, one obtains Eq.~\eqref{eq:dil_WI_gamma}. Proceeding along similar lines, 
one obtains the corresponding modified SCT WI for $\gamma$. For that one must use that
\begin{align}
\updelta_{\rm conf} R_k(x,y)&=(K_{\mu}^x+K_{\mu}^y-2(d-D_{\varphi})(x_{\mu}+y_{\mu}) \nonumber\\
&=-(x_{\mu}+y_{\mu})\partial_t R_k(x,y),
\end{align}
where the second equation is valid for regulator parametrized as in Eq.~\eqref{eq:regProfile}. From this one deduces:
\begin{align}
    \int_x \frac{\delta\gamma[\sigma]}{\delta\sigma(x)}&(K_{\mu}^x-2x_{\mu}(2D_{\varphi}-d))\sigma(x) \nonumber\\
    &=-\frac N2\int_{x,y}G[x,y](x_{\mu}+y_{\mu})\partial_t R_k(x,y).
    \label{eq:confgammaapp}
\end{align}

\section{Proof of Eq.~\eqref{eq:tadpole} and Eq.~\eqref{eq:id_conf} \label{app:identities}}

We first show a number of useful identities for sufficiently well-behaved functions that vanish at infinity, and then prove Eq.~\eqref{eq:tadpole} and Eq.~\eqref{eq:id_conf}. Importantly, the proof relies on the fact that in large $N$, $\eta=0$ and thus $D_\varphi=\frac{d-2}{2}$.

\begin{widetext}
\subsection{Useful identities}

It is useful to note that $\int_x f(x)D^x g(x)=\int_x g(x)(-d-D^x )f(x) $. Furthermore, one can easily show that $[\partial^2,D^x]=2\partial^2$, with $\partial^2=\partial_\nu\partial^\nu$. This implies the first identity
\begin{equation}\begin{split}
I_1&\equiv\int_{x} f(x)\partial^2 (D^x f( x)),\\
&=\int_{ x} f( x)\,\bigl[2+D^x]\,\partial^2 f( x),\\
&=\int_{ x}  \,\bigl[(2-d-D^x\bigr)f(x)]\,\partial^2 f( x),\\
&=(2-d)\int_{ x} f(x)\partial^2 f( x)-I_1,
\end{split}\end{equation}
and thus
\begin{equation}
    I_1 = -\frac{d-2}{2}\int_{ x} f(x)\partial^2 f( x).
\end{equation}
This, in turn, directly implies that 
\begin{equation}\begin{split}
\label{eq:id1}
\int_{ x} f( x)\partial^2\bigl[D_\varphi+D^x\bigr]f( x)=0.
\end{split}\end{equation}

Applying the same integration by parts strategy allows us to show 
\begin{equation}\begin{split}
I_2&\equiv\int_{ x} f( x)\sigma( x)\bigl[2-d-2 D^x\bigr]f( x),\\
&=\int_{ x} f( x)\,\bigl[2+d+ 2D^x\bigr]\!\bigl(\sigma( x)f( x)\bigr),\\
&=\int_{ x} f( x)\sigma( x)\bigl[2+d+2 D^x\bigr]f( x)
+\int_{ x} f( x)^2\,2 D^x\sigma( x),\\
&=-I_2+2\int_{ x} f( x)\,\bigl[(2+ D^x)\sigma( x)\bigr]\,f( x),
\end{split}\end{equation}
and thus
\begin{equation}
\label{id2}
    \int_{ x} f( x)\sigma( x)\bigl[2D_\varphi+2 D^x\bigr]f( x) = -\int_{ x} f( x)\,\bigl[(d-2D_\varphi+ D^x)\sigma( x)\bigr]\,f( x).
\end{equation}

For the SCT transformation, it is useful to note the following integration-by-parts property
\begin{equation}
\int_x g(x) \, K_\mu^x f(x)
= \int_x  f(x)\,\bigl(-K_\mu^x+2dx_\mu \bigr)g(x).
\end{equation}

We now show another identity, starting from
\begin{equation}
\begin{split}
I_3
&\equiv\int_x f(x) \, \sigma(x)\,\Bigl(2K_\mu^x-2(d-2)x_\mu\Bigr) f(x),\\
&= \int_x \Bigl(-2K_\mu^x\bigl(f(x)\sigma(x)\bigr) + 2(d+2)x_\mu\,f(x)\sigma(x)\Bigr)\, f(x)\\
&=-I_3 -2 \int_x f(x)^2\,\Bigl(K_\mu^x-4x_\mu\Bigr)\sigma(x),
\end{split}
\end{equation}
and thus
\begin{equation}
\begin{split}\label{eq:id3}
\int_x f(x) \, \sigma(x)\,\Bigl(2K_\mu^x-4D_\varphi x_\mu\Bigr) f(x)=- \int_x f(x)^2\,\Bigl(K_\mu^x-2(d-2D_\varphi)x_\mu\Bigr)\sigma(x).
\end{split}
\end{equation}

One final useful identity is $[\partial^2,K_\mu]=2(d-2)\partial_\mu-4x_\mu\partial^2$, which leads to
\begin{equation}
    \begin{split}
       I_4&\equiv \int_x f(x)\partial^2(K_\mu f(x)),\\
       &=  \int_x f(x)K_\mu \partial^2f(x)-4\int_x f(x)x_\mu \partial^2 f(x)+(d-2)\int_z\partial_\mu(f(z)^2),\\
       &=-I_4+2(d-2)\int_x f(x) \partial^2(x_\mu f(x))+(d-2)\int_x\partial_\mu(f(x)^2).
    \end{split}
\end{equation}
The last term on the right-hand side of the last line is the integral of a total derivative and thus vanishes, which implies 
\begin{equation}
    \begin{split}
       I_4= (d-2)\int_x f(x) \partial^2(x_\mu f(x)),
    \end{split}
\end{equation}
and in turn 
\begin{equation}
    \begin{split}\label{eq:id4}
       \int_x f(x)\partial^2([K_\mu-2D_\varphi x_\mu] f(x))=0.
    \end{split}
\end{equation}

\subsection{Proof of Eq.~\eqref{eq:tadpole}}
 With this preamble, we are now in a position to show that
\begin{equation}
    \int_{y,z} G[x,y] \, \partial_t R_k(y,z) \, G[z,x]=-\mathcal D
\bigl(G[x,x]-G_{0}(x,x)\bigr).
\end{equation}

To this end, we start from
\begin{equation}
\begin{split}
    T_1 \equiv  \int_{y,z} G[x,y] \, \partial_t R_k(y,z) \, G[z,x]&=\int_{y,z} G[x,y]\bigl(2d-2D_\varphi+D^y+D^z\bigr)\,
R_k(y,z)\,G[z,x],\\
&=-\int_{ y, z}\, G[ x, y]\,R_k( y, z)\,\bigl(2D_\varphi+2 D^z\bigr)\,
G[ z, x].
\end{split}
\end{equation}
Replacing $R_k( y, z)$ by $G^{-1}[ y, z]-\delta( y- z)(-\partial^2_{ z}+\sigma( z))$, we obtain
\begin{equation}
\begin{split}
    T_1 =& -\int_{ z}\! \delta( x- z)\,\bigl(2D_\varphi+2 D^z\bigr)\,
\left(G[ z, x]-G_{0}( z, x)\right)+\int_{ z}\, G[ x, z]\,[-\partial^2_{ z}+\sigma( z)]\,\bigl[2D_\varphi+2 D^z\bigr]\,
G[ z, x],\\
=&-\bigl[2D_\varphi+ D^x\bigr]\,
\left(G[ x, x]-G_{0}( x, x)\right)+\int_{ z}\, G[ x, z]\,[-\partial^2_{ z}+\sigma( z)]\,\bigl[2D_\varphi+2 D^z\bigr]\,
G[ z, x].
\end{split}
\end{equation}
We have added a zero term since $\bigl[2D_\varphi+ D^z+D^x\bigr]\,
G_{0}(z, x) = 0$ and thus by symmetry $\int_{ z}\! \delta(x- z)\,\bigl[2D_\varphi+2 D^z\bigr]\,
G_{0}(z, x)=0$. The subtraction is necessary to ensure UV convergence. Then we have used that $D^x$ acts on both arguments of $G[ x, x]$, and thus $\int_{ z}\! \delta( x- z) D^z\,
G[ z, x]=\frac12D^x G[ x, x]$, and similarly for $G_0(x,x)$

Using Eqs.~\eqref{eq:id1} and~\eqref{id2}, we have
\begin{equation}
\begin{split}
\int_{ z}\, G[ x, z]\,(-\partial^2_{ z}+\sigma( z))\,\bigl(2D_\varphi+2 D^z\bigr)\,
G[ z, x]&=-\int_{ z} G[ x, z]\,\bigl[(d-2D_\varphi+ D^z)\sigma( z)\bigr]\,G[ x, z] ,\\
&=\int_{ z}(d-2D_\varphi+ D^z)\sigma( z)\frac{\delta G[ x, x]}{\delta \sigma( z)},\\
&=\int_{ z}(d-2D_\varphi+ D^z)\sigma( z)\frac{\delta }{\delta \sigma( z)}\left(G[ x, x]-G_{0}( x, x)\right).
\end{split}
\end{equation}
Combining all these results together gives Eq.~\eqref{eq:tadpole}.

\subsection{Proof of  Eq.~\eqref{eq:id_conf}}
Finally, we wish to prove that
\begin{equation}
    \int_{y,z}\,G[x,y]\,\Bigl( K_\mu^y + K_\mu^z - (2d-2D_\varphi)(y_\mu+z_\mu)\Bigr)R_k(y,z)\,G[z,x]=-\mathcal{K}_\mu\Bigl(G[x,x]-G_0(x,x)\Bigr).
\end{equation}
The proof is similar to that just above.

We start from
\begin{equation}
\begin{split}
 T_2&=-\int_{y,z}\,G[x,y]\,\Bigl( K_\mu^y + K_\mu^z - 2(2d-2D_\varphi)(y_\mu+z_\mu)\Bigr)R_k(y,z)\,G[z,x],\\
&= -\int_{y,z}\,G[x,y]\,R_k(y,z)\,\Bigl(-2K_\mu^z+4D_\varphi z_\mu\Bigr)\,G[z,x],
\\
&= -\int_{y,z}\,G[x,y]\,\Bigl(G^{-1}[y,z]+\delta(y-z)\partial_z^2-\delta(y-z)\sigma(z)\Bigr)\,
\Bigl(-2K_\mu^z+4D_\varphi z_\mu\Bigr)\,G[z,x],\\
&=\int_{z}\delta(x-z)\Bigl(2K_\mu^z-4D_\varphi z_\mu\Bigr)\,(G[z,x]-G_0(z,x))-\int_{z}\,G[x,z]\,\Bigl(\partial_z^2-\sigma(z)\Bigr)\,\Bigl(-2K_\mu^z+4D_\varphi z_\mu\Bigr)\,G[z,x],\\
&=\Bigl(K_\mu^x-4D_\varphi x_\mu\Bigr)\,(G[x,x]-G_0(x,x))-\int_{z}\,G[x,z]\,\Bigl(\partial_z^2-\sigma(z)\Bigr)\,\Bigl(-2K_\mu^z+4D_\varphi z_\mu\Bigr)\,G[z,x],
\end{split}
\end{equation}
where we have used that $(K_\mu^x+K_\mu^z)G_0(z,x)=2D_\varphi (x^\mu+z^\mu)G_0(z,x)$ to subtract the term containing $G_0(z,x)$ on the fourth line.

Using Eqs.~\eqref{eq:id3} and~\eqref{eq:id4}, $T_2$ can be simplified,
\begin{equation}
    T_2=\mathcal{K}_\mu\Bigl(G[x,x]-G_0(x,x)\Bigr),
\end{equation}
proving Eq.~\eqref{eq:id_conf}.

\end{widetext}

\section{The independence in channels of the dilation WI}
\label{app:indepChanDil}

In this appendix, we shall demonstrate that the RHS of Eq.~\eqref{eq:DilEq} can be readily rewritten as the $t-$derivative of the sum of $I^{(n)}$'s that constitute $\gamma^{(n)}$. In order to do so, it is enough to consider only the diagrams of a particular channel and see that one obtains the $t-$derivative of the corresponding $I^{(n)}$. 

The procedure is straightforward once one notices that:
\begin{equation}
	\label{eq:derTProp}
	\partial_t G(q)=-\partial_t R_k(q^2)G^2(q).
\end{equation}
Each set of diagrams $H_i^{(n)}$ includes $n$ terms, which can be identified with the $\partial_t$ operator acting on each of the $n$ propagators that appear in the definition of $I^{(n)}$. For example,
\begin{equation}
	\begin{split}
		\int_q\partial_t R_k(q^2)G^2(q)&\prod_{i=1}^{n-1}G(q+\sum_{j=1}^{i}p_j)
		\\
		&= -\int_q \partial_tG(q)\prod_{i=1}^{n-1}G(q+\sum_{j=1}^{i}p_j).
	\end{split}
\end{equation}Thus, considering the sum of all the diagrams in a particular channel, one can rewrite (using the fact that the sum of derivatives is the derivative of the sum):
\begin{equation}
	\begin{split}
		\int_q \partial_t R_k(q^2)G^2(q)H^{(n)}_{j}\Big|_{q'=-q}=-\partial_t I^{(n)}\big(p_{P_j(1)},\dots,p_{P_j(n-1)}\big)
	\end{split},
\end{equation}
and repeat this process for each set of diagrams corresponding to a different channel. As a consequence, both the LHS and the RHS of Eq.~\eqref{eq:DilEq} are but linear operators acting on a linear combination of $I^{(n)}$'s, and thus instead of a single regularized dilation WI for each $\gamma^{(n)}$, we have $(n-1)!/2$ independent identities for every possible $I^{(n)}$.

\section{The $\partial_{r_\mu}\gamma^{(n+1)}$ and $(\partial_{q_\mu}+\partial_{q'_\mu})H^{(n)}_j$ terms in the SCT}\label{app:drmuandqmuH}

Let us stress that the analysis in the present Appendix is done for the particular channel configuration in which all momenta appear in the order $p_1, p_2, \dots, p_n$ (denoted ``channel 1'' in the main text).

In this section, it will prove useful to recall the definitions for functions $I^{(n)}$, $I_j^{(n+1)}$, $J_i^{(n)}$, and $J_{i,\mu}^{(n)}$:
\begin{widetext}

    \begin{equation}
		I^{(n)}(p_1,\dots,p_{n-1})\equiv\int_q \prod_{j=0}^{n-1}G\Big(Q_j\Big),
	\end{equation}

    \begin{equation}
    \begin{split}
        I^{(n+1)}_{i}(r;p_1,\dots,p_{n-1})\equiv &I^{(n+1)}(p_1,\dots,p_i,r,p_{i+1},\dots,p_{n-1})\\&=\int_q G(q)G(q+r)\Big\lbrace\prod_{j=1}^{n-i-1}G\Big(q+r+\sum_{l=1}^{j}p_{i+l}\Big)\Big\rbrace\Big\lbrace\prod_{j=1}^{i}G\Big(q'+\sum_{l=1}^{j}p_{i-l+1}\Big)\Big\rbrace_{q'=-q},
    \end{split}
	\end{equation}
    \begin{equation}
		J^{(n)}_{i}(p_1,\dots,p_{n-1})\equiv\int_q \partial_tR_k(q^2)G(q)^2\Big\lbrace\prod_{j=1}^{n-i-1}G\Big(q+\sum_{l=1}^{j}p_{i+l}\Big)\Big\rbrace\Big\lbrace\prod_{j=1}^{i}G\Big(q'+\sum_{l=1}^{j}p_{i-l+1}\Big)\Big\rbrace_{q'=-q},
	\end{equation}
	\begin{equation}
		J^{(n)}_{i,\mu}(p_1,\dots,p_{n-1})\equiv\int_q \partial_tR_k(q^2)G^2(q)\big(\partial _{q_{\mu}}+\partial _{q'_{\mu}}\big)\Big\lbrace\prod_{j=1}^{n-i-1}G\Big(q+\sum_{l=1}^{j}p_{i+l}\Big)\Big\rbrace\Big\lbrace\prod_{j=1}^{i}G\Big(q'+\sum_{l=1}^{j}p_{i-l+1}\Big)\Big\rbrace_{q'=-q},
	\end{equation}
\end{widetext}
with the index $i$ ranging between 0 and $n-1$, $Q_l=q+\sum_{j=1}^l p_j$ and $Q_0=q$.

\subsection{The left-hand side term $\partial_{r_\mu}I^{(n+1)}$}
When computing the $\partial_{r_\mu}$ contribution, in general, we have three different possibilities. In the first case, the $r$ momentum is at the beginning:
\begin{equation}
	\begin{split}    
		\label{eq:derivadaP1}
		\partial_{r_{\mu}}I^{(n+1)}
		(r,p_1,\dots,p_{n-1})&|_{r=0}=
		\\
		\partial_{r_{\mu}}\int_q G(q)G(q+r)\prod_{i=1}^{n-1}&G\big(q+r+\sum_{j=1}^{i}p_j\big)\Big|_{r=0} \\
		=\partial_{1\mu}\int_q G(q)^2 \prod_{i=1}^{n-1}G\big(&q+r+\sum_{j=1}^{i}p_j\big),
		\\
		+\frac{1}{2}\int_q\partial_{q_{\mu}}\big(G^2(q)\big)&\prod_{i=1}^{n-1}G\big(q+r+\sum_{j=1}^{i}p_j\big),\\
		=\frac{1}{2}\partial_{1\mu}I^{(n+1)}(0,p_1,\cdots&,p_{n-1})=\frac{1}{2}\partial_{1\mu}I^{(n+1)}_0\Big|_{r=0}.
	\end{split}
\end{equation}
In the last line of \eqref{eq:derivadaP1}, we have performed an integration by parts. Since both $q$ and $p_1$ appear in every propagator being differentiated after integrating by parts, the result is a total derivative $-\frac{1}{2}\partial_{1\mu}$, which changes the global coefficient from $1$ to $1/2$.

The second case is when the $r^\mu$ momentum is at the end of the permutation. A similar manipulation shows that
\begin{equation}
	\begin{split}    
		\label{eq:derivadaPnm1}
		&\partial_{r_{\mu}}I^{(n+1)}
		(p_1,\dots,p_{n-1},r)\Big|_{r=0}=
		\\
		&\partial_{r_{\mu}}\int_q G(q)G(q+r+\sum_{j=1}^{n-1}p_j)\prod_{i=1}^{n-1}G\big(q+\sum_{j=1}^{i}p_j\big)\Big|_{r=0},\\
		&=\frac{1}{2}\partial_{n-1\mu}I^{(n+1)}(p_1,\cdots,p_{n-1},0)=\frac{1}{2}\partial_{n-1\mu}I^{(n+1)}_{n-1}\Big|_{r=0}.
	\end{split}
\end{equation}

The third, and more complicated case, is when the $r^{\mu}$ momentum is somewhere in the middle. In particular, consider $r$ between $p_i$ and $p_{i+1}$. A more involved, yet similar, manipulation shows that this can be rewritten as:
\begin{equation}\label{eq:derivadaP3}
	\begin{split}
		\partial_{r_\mu}I^{(n+1)}(\dots,p_i,r,p_{i+1},\dots)\Big|_{r=0}&=\partial_{r_\mu}I^{(n+1)}_i\Big|_{r=0},\\
		&=\frac{1}{2}\left(\partial_{i\mu}+\partial_{i+1\mu}\right)I^{(n+1)}_i\Big|_{r=0}.
	\end{split}
\end{equation}

The addition of all three contributions yields:
\begin{equation}
\label{eq:A4}
	\begin{split}
		&\partial_{r_\mu}\sum_{j=0}^{n-1}I^{(n+1)}_j\Big|_{r=0}=\sum_{i=1}^{n-2}\frac{1}{2}\big(\partial_{i\mu}+\partial_{i+1\mu}\big)I^{(n+1)}_i\Big|_{r=0}\\
		&+\frac{1}{2}\Big(\partial_{1\mu}I^{(n+1)}_0+\partial_{n-1\mu}I^{(n+1)}_{n-1}\Big)\Big|_{r=0}\\
		&=\frac{1}{2}\sum_{i=1}^{n-1}\Big(\partial_{i\mu}I^{(n+1)}_i+\partial_{i\mu}I^{(n+1)}_{i-1}\Big)\Big|_{r=0}.
	\end{split}
\end{equation}
To go further, notice that a derivative with respect to $\sigma$ of $I^{(n)}$ gives rise to $n$ terms, since there are $n$ propagators and, in each one, one can apply the formula:
\begin{equation}\label{eq:derPropSigma}
	\partial_\sigma G(q;\sigma)=-G^2(q;\sigma).
\end{equation}
As a consequence, one obtains an expression of the derivative of $I^{(n)}$:
\begin{equation}
\label{eq:A6}
	\partial_{\sigma}I^{(n)}=-\sum_{j=0}^{n-1} I^{(n+1)}_j\Big|_{r=0}.
\end{equation}
When comparing Eqs.~\eqref{eq:A4} and an appropriate derivative of \eqref{eq:A6} one observes that they are very similar, except for some extra terms that we now make explicit:
\begin{equation}
	\begin{split}
		\partial_{r_{\mu}}&\sum_{j=0}^{n-1} I^{(n+1)}_j\Big|_{r=0}=
		-\frac{1}{2}\big(\sum_{i=1}^{n-1}\partial_{i\mu}\big)\partial_{\sigma}I^{(n)}\\
		-&\frac{1}{2}\sum_{i=1}^{n-2}\sum_{j=i+1}^{n-1}\Big[\partial_{i\mu}I^{(n+1)}_j+\partial_{j\mu}I^{(n+1)}_{i-1}\Big]\Big|_{r=0}.
	\end{split}
\end{equation}

\subsection{The right-hand side term $(\partial_{q_\mu}+\partial_{q'_\mu})H^{(n)}_j$}

To evaluate these contributions we start by considering the cases where the \textit{implicit} external leg $p_n$ is right next to the regulator insertion in the diagrams. There are two such contributions which correspond to $J^{(n)}_{0,\mu}$ and $J^{(n)}_{n-1,\mu}$. These are
\begin{equation}
	\begin{split}
		J^{(n)}_{0,\mu}=\int_q& \partial_t R_k(q^2)G^2(q)\\\times\big(\partial_{q_{\mu}}&+\partial_{q'_{\mu}}\big)\Big\lbrace\prod_{i=1}^{n-1}G\big(q+\sum_{j=1}^{i}p_j\big)\Big\rbrace\Big|_{q'=-q}\\&=\partial_{1\mu}J^{(n)}_{0},
	\end{split}
\end{equation}
and
\begin{equation}
	\begin{split}
		J^{(n)}_{n-1,\mu}=\int_q& \partial_t R_k(q^2)G^2(q)\\
		\times\big(\partial_{q_{\mu}}&+\partial_{q'_{\mu}}\big)\Big\lbrace\prod_{i=1}^{n-1}G\big(q'+\sum_{j=1}^{i}p_{n-j}\big)\Big\rbrace\Big|_{q'=-q}\\&=\partial_{n-1\mu}J^{(n)}_{n-1},
	\end{split}
\end{equation}
respectively.

As for the \textit{middle} terms, a treatment similar (and simpler) to the one exposed earlier in \eqref{eq:derivadaP3} holds yielding:
\begin{equation}
	J^{(n)}_{i,\mu}=\big(\partial_{i\mu}+\partial_{i+1\mu}\big)J^{(n)}_{i}.
\end{equation}
Again, these contributions can be rearranged, recalling Eq.~\eqref{eq:derTProp}. As explained in App.~\ref{app:indepChanDil}, the $\partial_t$ derivative of $I^{(n)}$ includes $n$ such terms -- one per each propagator on which the $t$ derivative acts on, and is given by:
\begin{equation}
	\partial_t I^{(n)}=-\sum_{j=0}^{n-1}J^{(n)}_{j}.
\end{equation}

Mimicking the procedure performed with the $\partial_{r_{\mu}}$, we notice that we almost have $(\sum\partial_{i\mu})$ of $\partial_t I^{(n)}$ in the right-hand side of the SCT identity Eq.~\eqref{eq:ChanSCTEq}, but some terms are missing. Adding and subtracting them allows us to write:

\begin{align}
	\sum_{j=0}^{n-1}J^{(n)}_{j,\mu}&=-\Big(\sum_i\partial_{i\mu}\Big)\partial_{t}I^{(n)}\nonumber\\
	&-\sum_{i=1}^{n-2}\sum_{j=i+1}^{n-1}\big[\partial_{i\mu}J^{(n)}_{j}+\partial_{j\mu}J^{(n)}_{i-1}\big].
\end{align}

\section{A final kinematic relation}\label{app:finalPart}

In this appendix, we show explicitly that the relation 
\begin{equation}\label{eq:combCDJ_app}
    \begin{split}
        \big[(n&-2)(2D_\varphi-d)\big]\sum_{i=1}^{n-1}\partial_{i\mu} I^{(n)} \\ 
        = &\sum_{i=1}^{n-2}\sum_{j=i+1}^{n-1}\Big[p_{j\nu} \partial_{i\mu} \partial_{j\nu} +p_{i\nu} \partial_{j\mu} \partial_{i\nu} -p_{j\mu}\partial_{j\nu} \partial_{i\nu} \\ 
        & -p_{i\mu}\partial_{i\nu} \partial_{j\nu} + p_{j\nu}\partial_{j\mu} \partial_{i\nu} + p_{i\nu}\partial_{i\mu} \partial_{j\nu} \Big] I^{(n)} \\ 
        & - \sum_{i=1}^{n-2}\sum_{j=i+1}^{n-1} \Big[ \partial_{i\mu}\big(J^{(n)}_j+(d-2D_\varphi)\sigma I^{(n+1)}_j\Big|_{r=0}\big) \\
        &+ \partial_{j\mu}\big(J^{(n)}_{i-1}+(d-2D_\varphi)\sigma I^{(n+1)}_{i-1}\Big|_{r=0}\big) \Big]
    \end{split}
\end{equation}
holds true. To show this, one notices that the propagator satisfies the following relation stemming from its fixed point equation: 
\begin{equation}\label{eq:appfpG}
    \partial_t G(q) + (d-2D_\varphi)\sigma \partial_\sigma G(q) + q_\nu \partial_{q_\nu} G(q) + 2G(q) = 0,
\end{equation}
which is straightforward to check once one uses the propagator and regulator profiles given in Eq.~\eqref{eq:propagator} and Eq.~\eqref{eq:regProfile}, respectively. Also, for this equation to hold, it is necessary to set $D_\varphi=\frac{d-2}{2}$ as is the case in the large N limit. 

The first step is to recognize that Eq.~\eqref{eq:appfpG} can be used to rewrite the combinations of the integrals ${J^{(n)}_j+(d-2D_\varphi)\sigma I^{(n+1)}_j}\Big|_{r=0}$ as
\begin{equation}
    \begin{split}
        &J^{(n)}_i+(d-2D_\varphi)\sigma I^{(n+1)}_i\Big|_{r=0} =\\ &\int_q \big(q_\nu \partial_{q_\nu} G(q) + 2G(q)\big) \Big\lbrace\prod_{j=1}^{n-i-1}G\Big(q+\sum_{l=1}^{j}p_{i+l}\Big)\Big\rbrace\\&\qquad\qquad \times\Big\lbrace\prod_{j=1}^{i}G\Big(q'+\sum_{l=1}^{j}p_{i-l+1}\Big)\Big\rbrace_{q'=-q}.
    \end{split}
\end{equation}

The first thing to notice is that the term proportional to $2G(q)$ when injected into Eq.~\eqref{eq:combCDJ_app} yields exactly:
\begin{equation}
    -2(n-2)\sum_{i=1}^{n-1}\partial_{i\mu} I^{(n)},
\end{equation}
in the right-hand side. This, in turn, cancels exactly the only left-hand side term once one uses $2D_\varphi-d=-2$. So, to complete the proof of equation \eqref{eq:combCDJ_app} we need to prove that:
\begin{equation}\label{eq:combCDJ_app2}
    \begin{split}
        &\sum_{i=1}^{n-2}\sum_{j=i+1}^{n-1}\Big[p_{j\nu} \partial_{i\mu} \partial_{j\nu} +p_{i\nu} \partial_{j\mu} \partial_{i\nu} -p_{j\mu}\partial_{j\nu} \partial_{i\nu} \\ 
        &\qquad -p_{i\mu}\partial_{i\nu} \partial_{j\nu} + p_{j\nu}\partial_{j\mu} \partial_{i\nu} + p_{i\nu}\partial_{i\mu} \partial_{j\nu} \Big] I^{(n)} \\ 
        &\qquad - \sum_{i=1}^{n-2}\sum_{j=i+1}^{n-1} \Big[ \partial_{i\mu}\big(L_{j}\big)+ \partial_{j\mu}\big(L_{i-1}\big) \Big]=0,
    \end{split}
\end{equation}
where we introduced the function:
\begin{equation}
    \begin{split}
        L_{i}\equiv&\int_qq_\nu \partial_{q_\nu}G(q) \Big\lbrace\prod_{j=1}^{n-i-1}G\Big(q+\sum_{l=1}^{j}p_{i+l}\Big)\Big\rbrace\\&\qquad\qquad \times\Big\lbrace\prod_{j=1}^{i}G\Big(q-\sum_{l=1}^{j}p_{i-l+1}\Big)\\=\int_q &\Big\lbrace\prod_{k=0,k\neq i}^{n-1}G(Q_k)\Big\rbrace\big[(Q_i)_\nu\partial_{i\nu}G(Q_i)\big],
    \end{split}
\end{equation}
where $Q_j=q+p_1+\dots+p_j$ and $Q_0=q$.

To manage the complexity of \eqref{eq:combCDJ_app2}, we first isolate and expand the following four-term derivative block from the first bracket. By distributing the derivatives, we obtain:

\begin{equation}
\begin{aligned}
    -\big(p_{i\mu}\partial_{i\nu}&\partial_{j\nu} - p_{i\nu}\partial_{j\nu}\partial_{i\mu} + p_{j\mu}\partial_{j\nu}\partial_{i\nu} - p_{j\nu}\partial_{i\nu}\partial_{j\mu}\big)I^{(n)} \\
    &= -\big(p_{i\mu}\partial_{j\nu}\partial_{i-1\nu} - p_{i\nu}\partial_{j\nu}\partial_{i-1\mu}\big)I^{(n)} \\
    + (p_{i\mu}\partial_{j\nu}&\delta_{\nu\rho} - p_{i\nu}\partial_{j\nu}\delta_{\mu\rho})\int_q \partial_{\rho}G(Q_{i-1})\prod_{k=0,k\neq i-1}^{n-1}G(Q_k) \\
    &- \big(p_{j\mu}\partial_{i\nu}\partial_{j+1\nu} - p_{j\nu}\partial_{i\nu}\partial_{j+1\mu}\big)I^{(n)} \\
    - (p_{j\mu}&\partial_{i\nu}\delta_{\nu\rho} - p_{j\nu}\partial_{i\nu}\delta_{\mu\rho})\int_q \partial_{\rho}G(Q_{j})\prod_{k=0,k\neq j}^{n-1}G(Q_k).
\end{aligned}
\end{equation}

Next, we process the first two lines in \eqref{eq:combCDJ_app2} as:
\begin{equation}\label{eq:auxReminder}
    \begin{split}
        \sum_{i=1}^{n-2}\sum_{j=i+1}^{n-1}&\left(\partial_{i\mu}p_{j\nu}\partial_{j\nu}+\partial_{j\mu}p_{i\nu}\partial_{i\nu}\right)I^{(n)} \\
        =\sum_{i=1}^{n-2}&\sum_{j=i+1}^{n-1}\left(\partial_{i\mu}p_{j\nu}\partial_{j+1\nu}+\partial_{j\mu}p_{i\nu}\partial_{i-1\nu}\right)I^{(n)}  
        \\&
        +\partial_{i\mu}\int_q p_{j\nu}\partial_{\nu}G(Q_{j})\prod_{k=0,k\neq j}^{n-1}G(Q_k)
        \\&-\partial_{j\mu}\int_q p_{i\nu}\partial_{\nu}G(Q_{i-1})\prod_{k=0,k\neq i-1}^{n-1}G(Q_k),
    \end{split}
\end{equation}
and combine them with the terms with $L_j$ and $L_{i-1}$ into:
\begin{equation}\label{eq:auxReminder2}
    \begin{split}
        \sum_{i=1}^{n-2}\sum_{j=i+1}^{n-1}&\left(\partial_{i\mu}p_{j\nu}\partial_{j\nu}+\partial_{j\mu}p_{i\nu}\partial_{i\nu}\right)I^{(n)} \\- &\sum_{i=1}^{n-2}\sum_{j=i+1}^{n-1} \Big[ \partial_{i\mu}\big(L_{j}\big)+ \partial_{j\mu}\big(L_{i-1}\big) \Big]
        =\\\sum_{i=1}^{n-2}&\sum_{j=i+1}^{n-1}\left(\partial_{i\mu}p_{j\nu}\partial_{j+1\nu}+\partial_{j\mu}p_{i\nu}\partial_{i-1\nu}\right)I^{(n)}  
        \\&
        -\partial_{i\mu}\int_q Q_{j-1\nu}\partial_{\nu}G(Q_{j})\prod_{k=0,k\neq j}^{n-1}G(Q_k)
        \\&-\partial_{j\mu}\int_q Q_{i\nu}\partial_{\nu}G(Q_{i-1})\prod_{k=0,k\neq i-1}^{n-1}G(Q_k).
    \end{split}
\end{equation}

By factoring the derivatives $\partial_{i-1\nu}$ and $\partial_{i\nu}$ out of the integral (since the contributions from the derivatives hitting these momenta add up to zero) in the last term and the derivatives $\partial_{j\nu}$ and $\partial_{j+1\nu}$ in the first to last, we have:
\begin{equation}
    \begin{split}
        \sum_{i=1}^{n-2}\sum_{j=i+1}^{n-1}&\left(\partial_{i\mu}p_{j\nu}\partial_{j\nu}+\partial_{j\mu}p_{i\nu}\partial_{i\nu}\right)I^{(n)} \\- &\sum_{i=1}^{n-2}\sum_{j=i+1}^{n-1} \Big[ \partial_{i\mu}\big(L_{j}\big)+ \partial_{j\mu}\big(L_{i-1}\big) \Big]
        =\\\sum_{i=1}^{n-2}&\sum_{j=i+1}^{n-1}\left(\partial_{i\mu}p_{j\nu}\partial_{j+1\nu}+\partial_{j\mu}p_{i\nu}\partial_{i-1\nu}\right)I^{(n)}  
        \\&
        -\partial_{i\mu}\partial_{j\nu}\int_q Q_{j-1\nu}\prod_{k=0}^{n-1}G(Q_k)\\&+\partial_{i\mu}\partial_{j+1\nu}\int_q Q_{j-1\nu}\prod_{k=0}^{n-1}G(Q_k)
        \\&-\partial_{j\mu}\partial_{i-1\nu}\int_q Q_{i\nu}\prod_{k=0}^{n-1}G(Q_k)\\&+\partial_{j\mu}\partial_{i\nu}\int_q Q_{i\nu}\prod_{k=0}^{n-1}G(Q_k).
    \end{split}
\end{equation}

Considering the terms with the cross partial derivatives in the $p_i$ and $p_j$ terms and moving the $\mu$ derivatives inside the integral one can write these two terms as:
\begin{widetext}
\begin{equation}
    \begin{split}
         \partial_{j\mu}&\partial_{i\nu}\int_q Q_{i\nu}\prod_{k=0}^{n-1}G(Q_k)-\partial_{i\mu}\partial_{j\nu}\int_q Q_{j-1\nu}\prod_{k=0}^{n-1}G(Q_k)=\\&\partial_{j+1\mu}\partial_{i\nu}\int_q Q_{i\nu} \prod_{k=0}^{n-1}G(Q_k)+\partial_{i\nu}\int_q Q_{j\nu}\partial_{\mu}G(Q_j)\prod_{\substack{k=0\\k\neq j}}^{n-1}G(Q_k)-\big(\sum_{l=i+1}^{j}p_{l\nu}\big)\partial_{i\nu}\int_q\partial_{\mu}G(Q_j)\prod_{\substack{k=0\\k\neq j}}^{n-1}G(Q_k)
         \\&-\partial_{i-1\mu}\partial_{j\nu}\int_q Q_{j-1\nu} \prod_{k=0}^{n-1}G(Q_k)+\partial_{j\nu}\int_q Q_{i-1\nu}\partial_{\mu}G(Q_{i-1})\prod_{\substack{k=0\\k\neq i-1}}^{n-1}G(Q_k)+\big(\sum_{l=i}^{j-1}p_{l\nu}\big)\partial_{j\nu}\int_q\partial_{\mu}G(Q_{i-1})\prod_{\substack{k=0\\k\neq i-1}}^{n-1}G(Q_k).
    \end{split}
\end{equation}
\end{widetext}
Now, all this construction was done to be able to perform, using the rotation WI for the propagator
\begin{equation}\label{eq:propRotWI}
    Q_\nu\partial_\mu G(Q)-Q_\mu\partial_\nu G(Q)=0,
\end{equation}
an exchange of indices $\mu\leftrightarrow\nu$ in the middle term of the last two lines. This will prove crucial to explicitly cancelling all terms. Performing this operation on these terms, we add everything together while reordering terms and taking the derivatives of propagators outside the integrals whenever simply possible (that is, when no explicit internal momentum is inside the integral), we have the following identity equivalent to Eq.~\eqref{eq:combCDJ_app2}:
\begin{widetext}
\begin{equation}\label{eq:comboCDJapp3}
\begin{split}
    \sum_{i=1}^{n-2}\sum_{j=i+1}^{n-1}&\bigg[p_{j\nu}\partial_{i\mu}\partial_{j+1\nu}I^{(n)} 
    + p_{i\nu}\partial_{j\nu}\partial_{i-1\mu}I^{(n)}
    +\big(\sum_{l=i}^{j-1}p_{l\nu}\big)\partial_{j\nu}(\partial_{i-1\mu}-\partial_{i\mu})I^{(n)}
    - p_{i\nu}\partial_{j\nu}(\partial_{i-1\mu}-\partial_{i\mu})I^{(n)}\\
    -&\partial_{i-1\mu}\partial_{j\nu}\int_q Q_{j-1\nu} \prod_{k=0}^{n-1}G(Q_k) 
    +\partial_{i\mu}\partial_{j+1\nu}\int_q Q_{j-1\nu}\prod_{k=0}^{n-1}G(Q_k)\bigg]\\
    +\sum_{i=1}^{n-2}\sum_{j=i+1}^{n-1}&\bigg[p_{i\nu}\partial_{j\mu}\partial_{i-1\nu}I^{(n)} 
    + p_{j\nu}\partial_{i\nu}\partial_{j+1\mu}I^{(n)}
    -\big(\sum_{l=i+1}^{j}p_{l\nu}\big)\partial_{i\nu}(\partial_{j\mu}-\partial_{j+1\mu})I^{(n)}
    + p_{j\nu}\partial_{i\nu}(\partial_{j\mu}-\partial_{j+1\mu})I^{(n)}\\&
    -\partial_{j\mu}\partial_{i-1\nu}\int_q Q_{i\nu}\prod_{k=0}^{n-1}G(Q_k)
    +\partial_{j+1\mu}\partial_{i\nu}\int_q Q_{i\nu} \prod_{k=0}^{n-1}G(Q_k)\bigg]\\
    +\sum_{i=1}^{n-2}\sum_{j=i+1}^{n-1}&\bigg[-p_{i\mu}\partial_{j\nu}\partial_{i-1\nu}I^{(n)} 
    -p_{j\mu}\partial_{i\nu}\partial_{j+1\nu}I^{(n)}
    +p_{i\mu}\partial_{j\nu}(\partial_{i-1\nu}-\partial_{i\nu})I^{(n)}
    -p_{j\mu}\partial_{i\nu}(\partial_{j\nu}-\partial_{j+1\nu})I^{(n)}\\
    &+\partial_{i\nu}\int_q Q_{j\mu}\partial_{\nu}G(Q_j)\prod_{\substack{k=0\\ k\neq j}}^{n-1}G(Q_k)
    +\partial_{j\nu}\int_q Q_{i-1\mu}\partial_{\nu}G(Q_{i-1})\prod_{\substack{k=0\\k\neq i-1}}^{n-1}G(Q_k)\bigg]=0.
\end{split}
\end{equation}
\end{widetext}

We have divided the expression into three structures which individually yield zero: i) a structure proportional to ``$\partial_{i\mu}\partial_{j\nu}$''; ii) a structure proportional to ``$\partial_{i\nu}\partial_{j\mu}$'' and iii) a structure proportional to ``$\partial_{i\nu}\partial_{j\nu}$''. All these terms can be worked around using shift in the indices. Indeed, since in these expressions the border terms involving $\partial_{0\rho}$ and $\partial_{n\rho}$ are 0 by construction, we can shift them by one. We have two such cases: a) $\partial_{j+1\rho}\to\partial_{j\rho}$ obtained by just adding and subtracting to the double sum a corresponding single summed term $\partial_{i+1\rho}$ and b) $\partial_{i-1\rho}\to\partial_{i\rho}$ obtained by just adding and subtracting to the double sum a corresponding single summed term $\partial_{j-1\rho}$.

Notice that the first and second bracketed expressions are almost identical with the exception that the roles of $j$ and $i$ are interchanged, while the shifts are in opposite directions for one and the other. We will work out explicitly the first bracketed expression, ``$\partial_{i\mu}\partial_{j\nu}$'', using this trick while the other one is entirely analogous.
\begin{widetext}
\begin{equation}
    \begin{split}
    \sum_{i=1}^{n-2}\sum_{j=i+1}^{n-1}&\bigg[p_{j\nu}\partial_{i\mu}\partial_{j+1\nu}I^{(n)} 
    + p_{i\nu}\partial_{j\nu}\partial_{i-1\mu}I^{(n)}
    +\big(\sum_{l=i}^{j-1}p_{l\nu}\big)\partial_{j\nu}(\partial_{i-1\mu}-\partial_{i\mu})I^{(n)}
    - p_{i\nu}\partial_{j\nu}(\partial_{i-1\mu}-\partial_{i\mu})I^{(n)}\\
    -&\partial_{i-1\mu}\partial_{j\nu}\int_q Q_{j-1\nu} \prod_{k=0}^{n-1}G(Q_k) 
    +\partial_{i\mu}\partial_{j+1\nu}\int_q Q_{j-1\nu}\prod_{k=0}^{n-1}G(Q_k)\bigg]=\\
    \sum_{i=1}^{n-2}\sum_{j=i+1}^{n-1}&\bigg[p_{j-1\nu}\partial_{i\mu}\partial_{j\nu}I^{(n)} 
    + p_{i\nu}\partial_{j\nu}\partial_{i\mu}I^{(n)}
    -\big(\sum_{l=i}^{j-1}p_{l\nu}\big)\partial_{j\nu}\partial_{i\mu}I^{(n)}
    +\big(\sum_{l=i+1}^{j-1}p_{l\nu}\big)\partial_{j\nu}\partial_{i\mu}I^{(n)}\\
    -&\partial_{i\mu}\partial_{j\nu}\int_q Q_{j-1\nu} \prod_{k=0}^{n-1}G(Q_k) 
    +\partial_{i\mu}\partial_{j\nu}\int_q Q_{j-2\nu}\prod_{k=0}^{n-1}G(Q_k)\bigg]\\
    -\sum_{i=1}^{n-1}&\bigg[p_{i\nu}\partial_{i\mu}\partial_{i+1\nu}I^{(n)}
    -\partial_{i\mu}\partial_{i+1\nu}\int_q Q_{i\nu} \prod_{k=0}^{n-1}G(Q_k) 
    +\partial_{i\mu}\partial_{i+1\nu}\int_q Q_{i-1\nu}\prod_{k=0}^{n-1}G(Q_k)\bigg]=\\
    \sum_{i=1}^{n-2}\sum_{j=i+1}^{n-1}&\bigg[p_{j-1\nu}\partial_{i\mu}\partial_{j\nu}I^{(n)}
    -\partial_{i\mu}\partial_{j\nu}\big(p_{j-1\nu}I^{(n)}\big)\bigg]
    +\sum_{i=1}^{n-1}\partial_{i+1\mu}I^{(n)}=0
    \end{split}
\end{equation}
\end{widetext}

Having proved that the first two brackets yield zero, it is only the last term that must be addressed. For this term the same type of manipulation must be performed.

\begin{widetext}
\begin{equation}
\begin{split}
    \sum_{i=1}^{n-2}\sum_{j=i+1}^{n-1}&\bigg[-p_{i\mu}\partial_{j\nu}\partial_{i-1\nu}I^{(n)} 
    -p_{j\mu}\partial_{i\nu}\partial_{j+1\nu}I^{(n)}
    +p_{i\mu}\partial_{j\nu}(\partial_{i-1\nu}-\partial_{i\nu})I^{(n)}
    -p_{j\mu}\partial_{i\nu}(\partial_{j\nu}-\partial_{j+1\nu})I^{(n)}\\
    &+\partial_{i\nu}\int_q Q_{j\mu}\partial_{\nu}G(Q_j)\prod_{\substack{k=0\\ k\neq j}}^{n-1}G(Q_k)
    +\partial_{j\nu}\int_q Q_{i-1\mu}\partial_{\nu}G(Q_{i-1})\prod_{\substack{k=0\\k\neq i-1}}^{n-1}G(Q_k)\bigg]=\\
    \sum_{i=1}^{n-2}\sum_{j=i+1}^{n-1}&\bigg[-p_{i\mu}\partial_{j\nu}\partial_{i\nu}I^{(n)}
    -p_{j\mu}\partial_{i\nu}\partial_{j\nu}I^{(n)}
    -\partial_{i\mu}I^{(n)}+\partial_{i\nu}(\partial_{j\nu}-\partial_{j+1\nu})\int_q Q_{j\mu}\prod_{k=0}^{n-1}G(Q_k)\\&
    -\partial_{j\mu}I^{(n)}+\partial_{j\nu}(\partial_{i-1\nu}-\partial_{i\nu})\int_q Q_{i-1\mu}\prod_{k=0}^{n-1}G(Q_k)\bigg]=\\
    \sum_{i=1}^{n-2}\sum_{j=i+1}^{n-1}&\bigg[-p_{i\mu}\partial_{j\nu}\partial_{i\nu}I^{(n)}
    -p_{j\mu}\partial_{i\nu}\partial_{j\nu}I^{(n)}
    -\partial_{i\mu}I^{(n)}+\partial_{i\nu}(\partial_{j\nu}-\partial_{j+1\nu})\int_q Q_{j\mu}\prod_{k=0}^{n-1}G(Q_k)\\&
    -\partial_{j\mu}I^{(n)}+\partial_{j\nu}(\partial_{i-1\nu}-\partial_{i\nu})\int_q Q_{i-1\mu}\prod_{k=0}^{n-1}G(Q_k)\bigg].
\end{split}
\end{equation}
\end{widetext}
Now, we need to regroup terms, in particular, those with explicit internal momentum inside the loop integral:
\begin{widetext}
\begin{equation}
\begin{split}
    \sum_{i=1}^{n-2}\sum_{j=i+1}^{n-1}&\bigg[-p_{i\mu}\partial_{j\nu}\partial_{i\nu}I^{(n)}
    -p_{j\mu}\partial_{i\nu}\partial_{j\nu}I^{(n)}
    -\partial_{i\mu}I^{(n)}+\partial_{i\nu}(\partial_{j\nu}-\partial_{j+1\nu})\int_q Q_{j\mu}\prod_{k=0}^{n-1}G(Q_k)\\&
    -\partial_{j\mu}I^{(n)}+\partial_{j\nu}(\partial_{i-1\nu}-\partial_{i\nu})\int_q Q_{i-1\mu}\prod_{k=0}^{n-1}G(Q_k)\bigg]=\\\sum_{i=1}^{n-2}\sum_{j=i+1}^{n-1}&\bigg[-p_{i\mu}\partial_{j\nu}\partial_{i\nu}I^{(n)}
    -p_{j\mu}\partial_{i\nu}\partial_{j\nu}I^{(n)}
    -\partial_{i\mu}I^{(n)}-\partial_{j\mu}I^{(n)}+\partial_{i\nu}\partial_{j\nu}\Big(\big(\sum_{l=i}^{j}p_{l\mu}\big)I^{(n)}\Big)\\&
    -\partial_{i\nu}\partial_{j+1\nu}\int_q Q_{j\mu}\prod_{k=0}^{n-1}G(Q_k)+\partial_{j\nu}\partial_{i-1\nu}\int_q Q_{i-1\mu}\prod_{k=0}^{n-1}G(Q_k)\bigg]=\\\sum_{i=1}^{n-2}\sum_{j=i+1}^{n-1}&\bigg[
    \big(\sum_{l=i+1}^{j-1}p_{l\mu}\big)\partial_{i\nu}\partial_{j\nu}I^{(n)}
    -\partial_{i\nu}\partial_{j\nu}\int_q Q_{j-1\mu}\prod_{k=0}^{n-1}G(Q_k)+\partial_{j\nu}\partial_{i\nu}\int_q Q_{i\mu}\prod_{k=0}^{n-1}G(Q_k)\bigg]\\+\sum_{i=1}^{n-1}&\bigg[\partial_{i\nu}\partial_{i+1\nu}\int_q Q_{i\mu}\prod_{k=0}^{n-1}G(Q_k)-\partial_{i\nu}\partial_{i+1\nu}\int_q Q_{i\mu}\prod_{k=0}^{n-1}G(Q_k)\bigg]=\\\sum_{i=1}^{n-2}\sum_{j=i+1}^{n-1}&\bigg[
    \big(\sum_{l=i+1}^{j-1}p_{l\mu}\big)\partial_{i\nu}\partial_{j\nu}I^{(n)}
    -\partial_{i\nu}\partial_{j\nu}\Big(\big(\sum_{l=i+1}^{j-1}p_{l\mu}\big)I^{(n)}\Big)\bigg]=0.
\end{split}
\end{equation}
\end{widetext}
which concludes the proof of Eq.~\eqref{eq:combCDJ_app} and consequently of Eq.~\eqref{eq:ChanSCTEq} implying the realization of conformal symmetry in the large N limit of $O(N)$ models.

\bibliography{ConfInvLargeN}

@article{Sonoda2017,
    author = {Sonoda, H.},
    title = {Conformal invariance for Wilson actions},
    journal = {Progress of Theoretical and Experimental Physics},
    volume = {2017},
    number = {8},
    pages = {083B05},
    year = {2017},
    month = {08},
    issn = {2050-3911},
    doi = {10.1093/ptep/ptx114},
    url = {https://doi.org/10.1093/ptep/ptx114},
    eprint = {https://academic.oup.com/ptep/article-pdf/2017/8/083B05/19650728/ptx114.pdf},
}

@article{Morris1997,
	title={Large N and the renormalization group},
	author={D'Attanasio, Marco and Morris, Tim R},
	journal={Phys. Lett. B},
	volume={409},
	number={1-4},
	pages={363--370},
	year={1997},
	publisher={Elsevier}
}

@article{Rosten:2017urs,
    author = "Rosten, Oliver J.",
    title = "{Wilsonian Ward Identities}",
    eprint = "1705.05837",
    archivePrefix = "arXiv",
    primaryClass = "hep-th",
    doi = "10.1140/epjc/s10052-019-6682-y",
    journal = "Eur. Phys. J. C",
    volume = "79",
    number = "2",
    pages = "176",
    year = "2019"
}

@article{Rosten:2016zap,
    author = "Rosten, Oliver J.",
    title = "{A Conformal Fixed-Point Equation for the Effective Average Action}",
    eprint = "1605.01729",
    archivePrefix = "arXiv",
    primaryClass = "hep-th",
    doi = "10.1142/S0217751X19500271",
    journal = "Int. J. Mod. Phys. A",
    volume = "34",
    number = "05",
    pages = "1950027",
    year = "2019"
}

@article{Rosten:2014oja,
    author = "Rosten, Oliver J.",
    title = "{On Functional Representations of the Conformal Algebra}",
    eprint = "1411.2603",
    archivePrefix = "arXiv",
    primaryClass = "hep-th",
    doi = "10.1140/epjc/s10052-017-5049-5",
    journal = "Eur. Phys. J. C",
    volume = "77",
    number = "7",
    pages = "477",
    year = "2017"
}

@article{Rosten:2016nmc,
    author = "Rosten, Oliver J.",
    title = "{A Wilsonian Energy-Momentum Tensor}",
    eprint = "1605.01055",
    archivePrefix = "arXiv",
    primaryClass = "hep-th",
    doi = "10.1140/epjc/s10052-018-5783-3",
    journal = "Eur. Phys. J. C",
    volume = "78",
    number = "4",
    pages = "312",
    year = "2018"
}

@article{Zamolodchikov1986,
	title={Irreversibility of the Flux of the Renormalization Group in a 2D Field Theory},
	author={Zamolodchikov, Alexander B},
	journal={JETP Lett.},
	volume={43},
	number={12},
	pages={730--732},
	year={1986}
}

@article{Polchinski1988,
	title = {Scale and conformal invariance in quantum field theory},
	journal = {Nucl. Phys. B},
	volume = {303},
	number = {2},
	pages = {226-236},
	year = {1988},
	issn = {0550-3213},
	doi = {https://doi.org/10.1016/0550-3213(88)90179-4},
	url = {https://www.sciencedirect.com/science/article/pii/0550321388901794},
	author = {Joseph Polchinski}
}

@mastersthesis{Cabrera2024,
  author       = {Cabrera, Santiago},
  title        = {Grupo de renormalización no perturbativo de los modelos $O(N)$: explorando la simetría conforme},
  school       = {Universidad de la República, Facultad de Ciencias (PEDECIBA)},
  year         = {2024},
  type         = {Master's thesis},
  address      = {Montevideo, Uruguay},
  url          = {https://hdl.handle.net/20.500.12008/45502}
}

@article{Cabrera2025,
	title = {Conformal invariance constraints in the $O(N)$ models: A study within the nonperturbative renormalization group},
	author = {Cabrera, Santiago and De Polsi, Gonzalo and Wschebor, Nicol\'as},
	journal = {Phys. Rev. E},
	volume = {111},
	issue = {5},
	pages = {054126},
	numpages = {18},
	year = {2025},
	month = {May},
	publisher = {American Physical Society},
	doi = {10.1103/PhysRevE.111.054126},
	url = {https://link.aps.org/doi/10.1103/PhysRevE.111.054126}
}

@article{Poland:2018epd,
    author = "Poland, David and Rychkov, Slava and Vichi, Alessandro",
    title = "{The Conformal Bootstrap: Theory, Numerical Techniques, and Applications}",
    doi = "10.1103/RevModPhys.91.015002",
    journal = "Rev. Mod. Phys.",
    volume = "91",
    pages = "015002",
    year = "2019"
}

@article{Delamotte2024,
	title = {Conformal invariance and composite operators: A strategy for improving the derivative expansion of the nonperturbative renormalization group},
	author = {Delamotte, Bertrand and De Polsi, Gonzalo and Tissier, Matthieu and Wschebor, Nicol\'as},
	journal = {Phys. Rev. E},
	volume = {109},
	issue = {6},
	pages = {064152},
	numpages = {18},
	year = {2024},
	month = {Jun},
	publisher = {American Physical Society},
	doi = {10.1103/PhysRevE.109.064152},
	url = {https://link.aps.org/doi/10.1103/PhysRevE.109.064152}
}

@article{DePolsi2019,
	author         = "De Polsi, Gonzalo and Tissier, Matthieu and Wschebor, Nicolás",
	title          = "{Conformal Invariance and Vector Operators in the O(N) Model}",
	journal        = "J. Stat. Phys.",
	volume         = "177",
	year           = "2019",
	pages          = "1089--1130",
	doi            = "10.1007/s10955-019-02411-3",
}

@Article{Pelissetto02,
	author   = {Andrea Pelissetto and Ettore Vicari},
	title    = {{Critical phenomena and renormalization-group theory}},
	journal  = {Physics Reports},
	year     = {2002},
	volume   = {368},
	pages    = {549 - 727},
	issn     = {0370-1573},
	doi      = {http://dx.doi.org/10.1016/S0370-1573(02)00219-3},
}

@article{Wilson:1973jj,
	author         = "Wilson, K. G. and Kogut, John B.",
	title          = "{The Renormalization group and the epsilon expansion}",
	journal        = "Phys. Rept.",
	volume         = "12",
	year           = "1974",
	pages          = "75-200",
	doi            = "10.1016/0370-1573(74)90023-4",
	SLACcitation   = "%%CITATION = PRPLC,12,75;%%"
}

@article{Pagani:2025dtc,
  title = {Explicit construction of the energy-momentum tensor in the large $N$ limit},
  author = {Pagani, Carlo and Sonoda, Hidenori},
  journal = {Phys. Rev. D},
  volume = {113},
  issue = {2},
  pages = {025012},
  numpages = {23},
  year = {2026},
  month = {Jan},
  publisher = {American Physical Society},
  doi = {10.1103/sfkr-vjkj},
  url = {https://link.aps.org/doi/10.1103/sfkr-vjkj}
}

@article{delamotte2016scale,
	title={Scale invariance implies conformal invariance for the three-dimensional Ising model},
	author={Delamotte, Bertrand and Tissier, Matthieu and Wschebor, Nicol{\'a}s},
	journal={Phys. Rev. E},
	volume={93},
	number={1},
	pages={012144},
	year={2016},
	publisher={APS}
}

@article{Delamotte:2007pf,
	author         = "Delamotte, Bertrand",
	title          = "{An Introduction to the nonperturbative renormalization
	group}",
	journal        = "Lect. Notes Phys.",
	volume         = "852",
	year           = "2012",
	pages          = "49-132",
	doi            = "10.1007/978-3-642-27320-9_2",
}

@article{Wetterich:1992yh,
	title = "Exact evolution equation for the effective potential",
	journal = "Phys. Lett. B",
	volume = "301",
	number = "1",
	pages = "90 - 94",
	year = "1993",
	issn = "0370-2693",
	doi = "https://doi.org/10.1016/0370-2693(93)90726-X",
	url = "http://www.sciencedirect.com/science/article/pii/037026939390726X",
	author = "Christof Wetterich"
}

@article{Ellwanger:1993kk,
	author         = "Ellwanger, U.",
	title          = "{Collective fields and flow equations}",
	journal        = "Z. Phys.",
	volume         = "C58",
	year           = "1993",
	pages          = "619-627",
	doi            = "10.1007/BF01553022",
	SLACcitation   = "%%CITATION = ZEPYA,C58,619;%%"
}

@article{Morris:1993qb,
	author         = "Morris, Tim R.",
	title          = "{The Exact renormalization group and approximate
	solutions}",
	journal        = "Int. J. Mod. Phys.",
	volume         = "A9",
	year           = "1994",
	pages          = "2411-2450",
	doi            = "10.1142/S0217751X94000972",
}

@article{Sonoda:2023ohb,
    author = "Sonoda, Hidenori",
    title = "{Exact Renormalization Group in Large $N$}",
    journal = "",
    eprint = "2302.09914",
    archivePrefix = "arXiv",
    month = "2",
    year = "2023"
}

@article{Sonoda:2015pva,
    author = "Sonoda, H.",
    title = "{Construction of the Energy-Momentum Tensor for Wilson Actions}",
    eprint = "1504.02831",
    archivePrefix = "arXiv",
    primaryClass = "hep-th",
    reportNumber = "KOBE-TH-15-03",
    doi = "10.1103/PhysRevD.92.065016",
    journal = "Phys. Rev. D",
    volume = "92",
    number = "6",
    pages = "065016",
    year = "2015"
}

@article{Polchinski:1983gv,
	author         = "Polchinski, Joseph",
	title          = "{Renormalization and Effective Lagrangians}",
	journal        = "Nucl. Phys.",
	volume         = "B231",
	year           = "1984",
	pages          = "269-295",
	doi            = "10.1016/0550-3213(84)90287-6",
	reportNumber   = "HUTP-83-A018",
	SLACcitation   = "%%CITATION = NUPHA,B231,269;%%"
}

@article{riva2005scale,
	title={Scale and conformal invariance in field theory: a physical counterexample},
	author={Riva, V and Cardy, J},
	journal={Phys. Lett. B},
	volume={622},
	number={3-4},
	pages={339--342},
	year={2005},
	publisher={Elsevier}
}

@book{ZinnJustin:2002ru,
  title={Quantum Field Theory and Critical Phenomena},
  author={Zinn-Justin, J.},
  year={2021},
  publisher={Oxford University Press}
}

@article{Dupuis:2020fhh,
    author = "Dupuis, N. and Canet, L. and Eichhorn, A. and Metzner, W. and Pawlowski, J. M. and Tissier, M. and Wschebor, N.",
    title = "{The nonperturbative functional renormalization group and its applications}",
    doi = "10.1016/j.physrep.2021.01.001",
    journal = "Phys. Rept.",
    volume = "910",
    pages = "1--114",
    year = "2021"
}

@article{Berges:2000ew,
	author         = "Berges, Juergen and Tetradis, Nikolaos and Wetterich,
	Christof",
	title          = "{Nonperturbative renormalization flow in quantum field
	theory and statistical physics}",
	journal        = "Phys. Rept.",
	volume         = "363",
	year           = "2002",
	pages          = "223-386",
}

@article{Moshe:2003xn,
	author         = "Moshe, Moshe and Zinn-Justin, Jean",
	title          = "{Quantum field theory in the large N limit: A Review}",
	journal        = "Phys. Rept.",
	volume         = "385",
	year           = "2003",
	pages          = "69-228",
	doi            = "10.1016/S0370-1573(03)00263-1",
}

@article{Morris:1994ie,
	author         = "Morris, Tim R.",
	title          = "{Derivative expansion of the exact renormalization
	group}",
	journal        = "Phys. Lett.",
	volume         = "B329",
	year           = "1994",
	pages          = "241-248",
	doi            = "10.1016/0370-2693(94)90767-6",
}

@article{Yabunaka:2018mju,
	author         = "Yabunaka, Shunsuke and Delamotte, Bertrand",
	title          = "{Why Might the Standard Large $N$ Analysis Fail in the
	O($N$) Model: The Role of Cusps in the Fixed Point
	Potentials}",
	journal        = "Phys. Rev. Lett.",
	volume         = "121",
	year           = "2018",
	number         = "23",
	pages          = "231601",
	doi            = "10.1103/PhysRevLett.121.231601",
}

@book{DiFrancesco:1997nk,
	author         = "Di Francesco, P. and Mathieu, P. and Senechal, D.",
	title          = "{Conformal Field Theory}",
	publisher      = "Springer-Verlag",
	address        = "New York",
	year           = "1997",
	url            = "http://www-spires.fnal.gov/spires/find/books/www?cl=QC174.52.C66D5::1997",
	series         = "Graduate Texts in Contemporary Physics",
	doi            = "10.1007/978-1-4612-2256-9",
	ISBN           = "9780387947853, 9781461274759",
	SLACcitation   = "%%CITATION = INSPIRE-454643;%%"
}

@article{Balog2020,
  title = {Conformal invariance in the nonperturbative renormalization group: A rationale for choosing the regulator},
  author = {Balog, Ivan and De Polsi, Gonzalo and Tissier, Matthieu and Wschebor, Nicol\'as},
  journal = {Phys. Rev. E},
  volume = {101},
  issue = {6},
  pages = {062146},
  numpages = {5},
  year = {2020},
  month = {Jun},
  publisher = {American Physical Society},
  doi = {10.1103/PhysRevE.101.062146},
  url = {https://link.aps.org/doi/10.1103/PhysRevE.101.062146}
}

@article{Blaizot2006,
author = {Blaizot, Jean Paul and M{\'{e}}ndez-Galain, Ram{\'{o}}n and Wschebor, Nicol{\'{a}}s},
doi = {10.1016/j.physletb.2005.10.086},
issn = {03702693},
journal = {Phys. Lett. B},
month = {jan},
number = {4},
pages = {571--578},
publisher = {North-Holland},
title = {{A new method to solve the non-perturbative renormalization group equations}},
volume = {632},
year = {2006}
}

@article{Yabunaka23,
  title = "{One Fixed Point Can Hide Another One: Nonperturbative Behavior of the Tetracritical Fixed Point of O($N$) Models at Large $N$}",
  shorttitle = {One Fixed Point Can Hide Another One},
  author = {Yabunaka, Shunsuke and Delamotte, Bertrand},
  year = 2023,
  month = jun,
  journal = {Phys. Rev. Lett.},
  volume = {130},
  number = {26},
  pages = {261602},
  publisher = {American Physical Society},
  doi = {10.1103/PhysRevLett.130.261602},
  urldate = {2026-04-12},
  langid = {english}
}

@article{Yabunaka17,
    title = {Surprises in {O}({N}) models: nonperturbative fixed points, large {N} limits, and multicriticality},
    volume = {119},
    doi = {10.1103/PhysRevLett.119.191602},
    number = {19},
    journal = {Phys. Rev. Lett.},
    publisher = {American Physical Society},
    author = {Yabunaka, Shunsuke and Delamotte, Bertrand},
    month = nov,
    year = {2017},
    pages = {191602},
}

@article{knorrExactSolutionsResidual2021,
    title = {Exact solutions and residual regulator dependence in functional renormalisation group flows},
    volume = {54},
    doi = {10.1088/1751-8121/ac00d4},
    number = {27},
    urldate = {2025-02-11},
    journal = {J. Phys. A},
    author = {Knorr, Benjamin},
    year = {2021},
    pages = {275401},
}

\end{document}